\documentclass[12pt,a4paper]{article}

\usepackage{amsmath,amssymb}
\usepackage[nosort]{cite}
\usepackage[hyperref,bulletsep]{collect}
\def\gfxon{\usepackage[final]{graphicx}}

\gfxon

\makeatletter \@addtoreset{equation}{section} \makeatother

\allowdisplaybreaks[3]

\newenvironment{bulletlist}{\begin{list}{$\bullet$}{\leftmargin1.5em\itemsep0pt}}{\end{list}}

\makeatletter
\let\old@startsection=\@startsection
\let\oldl@section=\l@section
\renewcommand{\@startsection}[6]{\old@startsection{#1}{#2}{#3}{#4}{#5}{#6\mathversion{bold}}}
\renewcommand{\l@section}[2]{%
\vspace{-0.5em}%
\oldl@section{\mathversion{bold}#1}{#2}}
\makeatother

\makeatletter
\let\old@makecaption=\@makecaption
\def\@makecaption{\small\old@makecaption}
\makeatother

\let\oldPhi=\Phi
\let\oldPsi=\Psi
\let\oldGamma=\Gamma
\let\oldDelta=\Delta
\let\oldSigma=\Sigma
\let\oldTheta=\Theta
\let\oldPi=\Pi
\renewcommand{\Phi}{\mathnormal{\oldPhi}}
\renewcommand{\Psi}{\mathnormal{\oldPsi}}
\renewcommand{\Gamma}{\mathnormal{\oldGamma}}
\renewcommand{\Sigma}{\mathnormal{\oldSigma}}
\renewcommand{\Delta}{\mathnormal{\oldDelta}}
\renewcommand{\Theta}{\mathnormal{\oldTheta}}
\renewcommand{\Pi}{\mathnormal{\oldPi}}

\newcommand{\ham}{\mathcal{H}}

\newcommand{\yang}{\mathcal{Y}}
\newcommand{\yangsing}{\mathcal{X}}
\newcommand{\shift}{\mathcal{U}}
\newcommand{\gen}[1]{\mathfrak{#1}}
\newcommand{\superN}{\mathcal{N}}

\newcommand{\cder}{\mathcal{D}}

\newcommand{\order}[1]{\mathcal{O}(#1)}

\newcommand{\Reals}{\mathbb{R}}

\ifx\genfrac\sdflkaj
\newcommand{\atopfrac}[2]{{{#1}\above0pt{#2}}}
\else
\newcommand{\atopfrac}[2]{\genfrac{}{}{0pt}{}{#1}{#2}}
\fi
\newcommand{\sfrac}[2]{{\textstyle\frac{#1}{#2}}}
\newcommand{\half}{\sfrac{1}{2}}
\newcommand{\ihalf}{\sfrac{i}{2}}

\newcommand{\indup}[1]{_{\mathrm{#1}}}

\newcommand{\rep}[1]{{\mathbf{#1}}}

\newcommand{\alg}[1]{\mathfrak{#1}}
\newcommand{\grp}[1]{\mathrm{#1}}

\newcommand{\MMM}[2]{{\arraycolsep0pt\begin{array}[b]{c}\makebox[0cm]{$\atopfrac{#2}{\downarrow}$}\\#1\end{array}}}

\newcommand{\lrbrk}[1]{\left(#1\right)}
\newcommand{\bigbrk}[1]{\bigl(#1\bigr)}

\newcommand{\bigcomm}[2]{\big[#1,#2\big]}
\newcommand{\comm}[2]{[#1,#2]}

\newcommand{\acomm}[2]{\{#1,#2\}}

\newcommand{\gcomm}[2]{[#1,#2\}}
\newcommand{\biggcomm}[2]{\big[#1,#2\big\}}

\newcommand{\state}[1]{\mathopen{|}#1\mathclose{\rangle}}
\newcommand{\bigstate}[1]{\bigl|#1\bigr\rangle}

\newcommand{\nn}{\nonumber}
\newcommand{\nln}{\nonumber\\} 
\newcommand{\nl}[1][0pt]{\nonumber\\[#1]&\hspace{-4\arraycolsep}&\mathord{}}
\newcommand{\nlnum}{\\&\hspace{-4\arraycolsep}&\mathord{}}
\newcommand{\earel}[1]{\mathrel{}&\hspace{-2\arraycolsep}#1\hspace{-2\arraycolsep}&\mathrel{}}
\newcommand{\eq}{\earel{=}}

\def\[{\begin{equation}}
\def\]{\end{equation}}
\def\<{\begin{eqnarray}}
\def\>{\end{eqnarray}}

\makeatletter
\def\mr@ignsp#1 {\ifx\:#1\@empty\else #1\expandafter\mr@ignsp\fi}%
\newcommand{\multiref}[1]{\begingroup
\xdef\mr@no@sparg{\expandafter\mr@ignsp#1 \: }%
\def\mr@comma{}%
\@for\mr@refs:=\mr@no@sparg\do{\mr@comma\def\mr@comma{,}\ref{\mr@refs}}%
\endgroup}
\makeatother

\newcommand{\hypref}[2]{\ifx\href\asklfhas #2\else\href{#1}{#2}\fi}

\newcommand{\secref}[1]{Sec.~\multiref{#1}}

\newcommand{\appref}[1]{App.~\multiref{#1}}

\newcommand{\figref}[1]{Fig.~\multiref{#1}}
\renewcommand{\eqref}[1]{(\multiref{#1})}

\ifx\href\asklfhas\newcommand{\href}[2]{#2}\fi
\newcommand{\arxivno}[1]{\href{http://arxiv.org/abs/#1}{#1}}

\begin{document}

\thispagestyle{empty}
\begin{flushright}\footnotesize
\texttt{\arxivno{arxiv:0707.1031}}\\
\texttt{AEI-2007-096}\\
\texttt{PUTP-2232}
\end{flushright}
\vspace{0cm}

\begin{center}%
{\Large\textbf{\mathversion{bold}%
On Symmetry Enhancement in the\\%
$\alg{psu}(1,1|2)$ Sector of $\superN = 4$ SYM}\par}\vspace{1cm}%

\textsc{Niklas~Beisert$^{1,2}$ and Benjamin~I.~Zwiebel$^{2}$} \vspace{8mm}

\textit{$^{1}$ Max-Planck-Institut f\"ur Gravitationsphysik\\%
Albert-Einstein-Institut\\%
Am M\"uhlenberg 1, 14476 Potsdam, Germany}%
\vspace{3mm}%

\textit{$^{2}$ Joseph Henry Laboratories\\%
Princeton University\\%
Princeton, NJ 08544, USA}%
\vspace{3mm}

\texttt{nbeisert@aei.mpg.de}\\
\texttt{bzwiebel@princeton.edu}\par\vspace{1cm}

\textbf{Abstract}\vspace{7mm}

\begin{minipage}{14.7cm}
Strong evidence indicates that the spectrum of planar anomalous dimensions 
of $\superN = 4$ super Yang-Mills theory 
is given asymptotically by Bethe equations. 
A curious observation is that the Bethe equations 
for the $\alg{psu}(1,1|2)$ subsector lead to very large degeneracies 
of $2^M$ multiplets, which apparently do not 
follow from conventional integrable structures. 
In this article, we explain such degeneracies 
by constructing suitable conserved nonlocal generators
acting on the spin chain.
We propose that they generate a
subalgebra of the loop algebra for the $\alg{su}(2)$ 
automorphism of $\alg{psu}(1,1|2)$.
Then the degenerate multiplets of size $2^M$ transform 
in irreducible tensor products of $M$ 
two-dimensional evaluation representations of the loop algebra. 
\end{minipage}

\end{center}

\newpage
\setcounter{page}{1}
\renewcommand{\thefootnote}{\arabic{footnote}}
\setcounter{footnote}{0}



\section{Introduction}

Methods of integrability have become a central tool for investigating 
the dynamics of planar $\superN=4$ extended supersymmetric gauge theory
and noninteracting strings on the $AdS_5\times S^5$ background
\cite{Minahan:2002ve,Beisert:2003tq,Bena:2003wd,Beisert:2003yb,Beisert:2005fw},
cf.~\cite{Beisert:2004ry,Plefka:2005bk} for reviews.
Investigations of the S-matrix
\cite{Staudacher:2004tk,Beisert:2005tm,Arutyunov:2006ak}
have recently led to a highly nontrivial test 
of the AdS/CFT correspondence 
showing that it may correctly interpolate
between weak and strong coupling 
\cite{Janik:2006dc,Beisert:2006ib,Bern:2006ew,Beisert:2006ez}.
The proposal has since been tested thoroughly,
see \cite{Benna:2006nd,Kotikov:2006ts,Maldacena:2006rv,Cachazo:2006az,
Alday:2007qf,Kostov:2007kx,Dorey:2007xn,Beccaria:2007tk,Gromov:2007cd,
Beccaria:2007cn,Kotikov:2007cy,Roiban:2007jf,Klose:2007rz,Beisert:2007hz,Casteill:2007ct, Chen:2007vs}.

Perturbative gauge theory in the planar limit can be cast into the
form of a spin chain. 
This spin chain model has a $\alg{psu}(2,2|4)$ symmetry,
and the spins transform in a noncompact module of the symmetry algebra.
At leading order this spin chain model
agrees with the standard nearest-neighbor integrable spin chain model 
based on this algebra and module \cite{Beisert:2003jj,Beisert:2003yb}.

Dealing with perturbative corrections
to the spin chain Hamiltonian and symmetry generators
is however a formidable problem:
With increasing order in perturbation theory the 
local interactions along the spin chain 
will act on more and more neighboring sites.
Moreover, higher-order interactions change the length
of the chain; they are dynamic \cite{Beisert:2003ys}. 
Together with the infinite degrees of freedom at each site,
the interactions become combinatorially almost intractable,
even at relatively low perturbative orders.
This holds true for obtaining them
(through explicit evaluation of gauge theory Feynman diagrams
or through clever construction)
as well as for applying them to states.
Furthermore, one can hardly rely on 
standard $\alg{psu}(2,2|4)$ representation theory 
because the algebra is not realized in a manifest way. 
Nevertheless, the commutation relations are essential in
constraining the form of the corrections.

As a step toward the complete corrections
at the first few loop orders
one can restrict to certain subsectors.
An apt choice is the $\alg{psu}(1,1|2)$
sector, which has complexity well balanced between
realistic features and simplifications.
It incorporates a noncompact spin representation
whose components are quite simple to enumerate.
Furthermore, the dynamic interactions are
mostly frozen out: The generators change the length
by a definite amount, either by one unit or not at all.
Finally, the Hamiltonian is a nonseparable part
of the symmetry algebra. 

The construction of the higher-loop algebra
for this sector was started in \cite{Zwiebel:2005er} (also see \cite{Belitsky:2005bu} for the two-loop dilatation 
generator of a $\alg{sl}(2)$ subsector).
A key simplification in this construction was
based on some less obvious symmetries:
In $\superN=4$ SYM the symmetry algebra of
the sector contains two factors of
$\alg{psu}(1|1)$ in addition to the
$\alg{psu}(1,1|2)$ algebra.
They made it possible to find the Hamiltonian at the two-loop level
and to represent it using simple building blocks.
Beyond that order, the construction appears to be rather complex. 
However, it might be that some crucial insight is still lacking 
in order to extend the construction conveniently to higher orders.

For example, a curious observation made in \cite{Beisert:2005fw}
has not yet been explained or taken into account: 
The Bethe equations for the sector lead to a huge
degeneracy of $2^M$ multiplets that is not explained by any known symmetries of the integral model. 
In this paper we would like to understand this 
degeneracy at the level of spin chain operators
commuting with the Hamiltonian. 
These might be of help in the construction of higher-loop corrections
to the algebra generators. 

The degeneracy is partially explained by an $\alg{su}(2)$ automorphism 
of the $\alg{psu}(1,1|2)$ algebra, see e.g.~\cite{Gotz:2005ka}.
The automorphism is not a part of the underlying $\alg{psu}(2,2|4)$ algebra
of $\superN=4$ SYM. It is nevertheless an exact symmetry
of the $\alg{psu}(1,1|2)$ sector, i.e.~it should apply also at finite $N\indup{c}$. 
The degenerate $\alg{psu}(1,1|2)$ multiplets 
transform in a tensor product of $\alg{su}(2)$ doublets, $\rep{2}^{\otimes M}$. 
However, such tensor products are reducible, and therefore 
the $\alg{su}(2)$ automorphism alone cannot explain the full degeneracy.

With respect to $\alg{su}(2)$, the multiplets transform in a reducible
$\rep{2}^{\otimes M}=\rep{2}\otimes\ldots\otimes \rep{2}$ representation.
This is reminiscent of the $\alg{su}(n)$ Haldane-Shastry model \cite{Haldane:1988gg,SriramShastry:1988gh},
which also has degenerate states transforming in reducible tensor products 
of $\alg{su}(n)$ representations \cite{Haldane:1992sj}.
There, the degeneracy is caused by a $\alg{su}(n)$ Yangian algebra
that commutes exactly with the Hamiltonian, even on a finite periodic chain. 
It is therefore conceivable that a $\alg{su}(2)$ Yangian 
or a similar algebraic structure 
will explain the further degeneracy in our case as well. 
In the present paper we shall present evidence in favor of this conjecture.

In Section \ref{sec:BetheSym}, we review the Bethe equations 
and transfer matrix and use them to observe this degeneracy. 
In Section \ref{sec:LieSym}, we review the leading-order 
spin representations for the $\alg{psu}(1,1|2)$ and $\alg{psu}(1|1)^2$ symmetry 
generators and present the $\alg{su}(2)$ automorphism. 
To gain further intuition about the degeneracy, 
we study some degenerate spin chain states in Section \ref{sec:states}. 
Finally, in Section \ref{Non-local} we explain the degeneracy by constructing 
an infinite set of nonlocal spin chain symmetry generators, at leading order. 
These generators are built from the $\alg{psu}(1|1)^2$ generators
and form a triplet of $\alg{su}(2)$.
We discuss how these new generators map between degenerate states 
and argue that they form a parabolic subalgebra of the loop algebra of $\alg{su}(2)$. 
We also discuss the relation of this symmetry to the integrable model's Yangian symmetry.  
Directions for further research are given in Section \ref{sec:conc}. 
Appendix \ref{sec:algebra} contains the commutation relations 
for the extended $\alg{psu}(1,1|2)$ and $\alg{psu}(1|1)^2$ algebras, 
and in Appendix \ref{app:multilinear} we present relevant multilinear 
operators for the $\alg{psu}(1,1|2)$ sector, 
including a cubic operator that is a $\alg{su}(2)$-triplet and $\alg{psu}(1,1|2)$ invariant. 
The proof that the nonlocal symmetry generators commute with the classical $\alg{psu}(1,1|2)$ generators 
and the one-loop dilatation generator is given in Appendix \ref{app:biproof}.

\section{Symmetry Enhancement in the Bethe Ansatz} \label{sec:BetheSym}

In this section, we describe the symmetries of the one-loop Bethe equations 
for the $\alg{psu}(1,1|2)$ sector, 
as well as the resulting $2^M$-fold degeneracies in the spectrum. 
Furthermore, we show that these degeneracies are also present for the transfer matrix, 
which provides the full set of local conserved charges of the integrable system. 

\subsection{Bethe Equations}

\begin{figure}\centering
\includegraphics{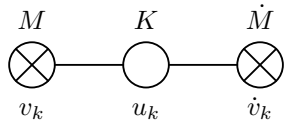}
\caption{Dynkin diagram for $\alg{psu}(1,1|2)$.
The different flavors of Bethe roots and their overall numbers are indicated
below/above the nodes, respectively.}
\label{fig:dynkin}
\end{figure}

The Bethe equations for the $\alg{psu}(1,1|2)$ sector of planar $\superN=4$ SYM
at leading order take the form
\<\label{eq:Bethe}
1\eq
\prod_{j=1}^K\frac{v_k-u_j-\ihalf}{v_k-u_j+\ihalf}\,,
\nln
1\eq
\lrbrk{\frac{u_k-\ihalf}{u_k+\ihalf}}^L
\mathop{\prod_{j=1}^K}_{j\neq k}\frac{u_k-u_j+i}{u_k-u_j-i}
\prod_{j=1}^{M}\frac{u_k-v_j-\ihalf}{u_k-v_j+\ihalf}
\prod_{j=1}^{\dot M}\frac{u_k-\dot v_j-\ihalf}{u_k-\dot v_j+\ihalf}\,,
\nln
1\eq
\prod_{j=1}^K\frac{\dot v_k-u_j-\ihalf}{\dot v_k-u_j+\ihalf}\,.
\>
These are just the standard Bethe equations
for a closed nearest-neighbor spin chain with 
$\alg{psu}(1,1|2)$ symmetry 
(in the form determined by the Dynkin diagram in \figref{fig:dynkin})
and spins transforming in the $[0;1;0]$ representation.
The three types of Bethe roots 
$v_{1,\ldots, M}$,
$u_{1,\ldots, K}$ and
$\dot v_{1,\ldots, \dot M}$
are associated to the three nodes of the Dynkin 
diagram in \figref{fig:dynkin}.
The length of the spin chain is given by $L$.

The momentum and energy eigenvalues for eigenstates of this system 
are determined through 
the main Bethe roots $u_{1,\ldots, K}$ alone
\[\label{eq:MomEng}
\exp(iP)=\prod_{j=1}^K \frac{u_j+\ihalf}{u_j-\ihalf}\,,
\qquad
E=\sum_{j=1}^K\lrbrk{\frac{2 i}{u_j+\ihalf}-\frac{2 i}{u_j-\ihalf}}.
\]
%

\subsection{Symmetries}
\label{sec:BetheSymText}

The $\alg{psu}(1,1|2)$ symmetry is realized in the standard way:
One can add Bethe roots $v,u,\dot v=\infty$ to the
set of Bethe roots for any eigenstate. 
It is easy to convince oneself 
that the Bethe equations \eqref{eq:Bethe} 
for the original roots as well as for the new root are satisfied. 
Moreover, the momentum and energy \eqref{eq:MomEng} are not changed by 
the introduction of the additional root. 
This means that the eigenstates come in highest-weight multiplets 
with degenerate momentum and energy eigenvalues. 
These multiplets are modules of the symmetry algebra 
$\alg{psu}(1,1|2)$.
Note that the Bethe roots $v,u,\dot v=\infty$
are allowed to appear in eigenstates more than one time, 
and thus even very large or infinite multiplets can
be swept out with this symmetry.

Another type of symmetry that is very important to
$\superN=4$ SYM exists only in the zero-momentum sector.
Here one adds a single root $v=0$ or $\dot v=0$ 
to an eigenstate configuration while decreasing
the length $L$ by one unit \cite{Beisert:2005fw}.
The original Bethe equations are preserved, and the
Bethe equation for $v=0$ and $\dot v=0$ is equal 
to the zero-momentum condition, cf.~\eqref{eq:MomEng}.
As the momentum and energy eigenvalues depend explicitly on the main
Bethe roots $u_k$ only, they are not affected by this transformation. 
This symmetry leads to an additional fourfold degeneracy 
of states because each of the Bethe roots $v=0$ and $\dot v=0$
can only appear once at maximum. 
The associated algebra consists of two copies of $\alg{su}(1|1)$ 
whose typical modules are two-dimensional.
These two additional algebras are required for a
consistent embedding of the spin chain into a 
larger model with $\alg{psu}(2,2|4)$ symmetry \cite{Beisert:2004ry}.
Their generators were constructed 
in \cite{Beisert:2004ry,Zwiebel:2005er}
at the leading order, and they transform one site of the spin chain 
into two or vice versa. We will present these generators in Section \ref{sec:LieSym}.

The third and most obscure type of symmetry was
observed in \cite{Beisert:2005fw}.
The auxiliary Bethe roots
$v_k$ and $\dot v_k$ appear in the Bethe equations
\eqref{eq:Bethe} completely symmetrically:
The Bethe equation for $v_k$ is exactly
the same as the one for $\dot v_k$. 
Furthermore, the product in the Bethe equation 
for $u_k$ involves a product over all $v_j$ 
and $\dot v_j$ with the same form of factor. 
Therefore, we can freely interchange them
\[\label{eq:SymBethe}
v_j\longleftrightarrow\dot v_{j'}
\]
without violating the Bethe equations.
As for the previous type of symmetry,
modifying only the auxiliary Bethe roots
does not change the momentum nor the energy.
It is straightforward to convince oneself that
this leads to a degeneracy of $2^{M_0}$ 
states where $M_0$ is the number of
$v_j$ roots which are distinct from $\dot v_j$
(in order to avoid coincident Bethe roots of the same type).

The closer investigation of this latter symmetry 
will be the main subject of the present paper.

\subsection{Commuting Charges}

\begin{figure}\centering
\parbox[c]{10cm}{\centering\includegraphics{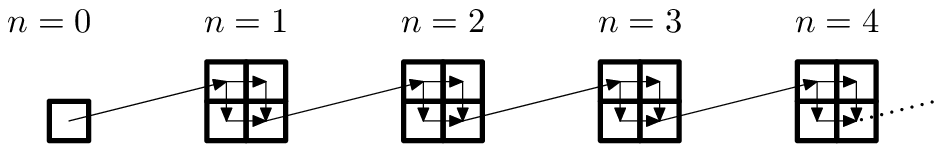}}
\qquad
\parbox[c]{3cm}{\centering\includegraphics{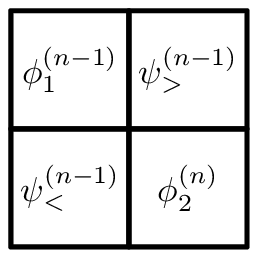}}
\caption{Structure of the spin representation (left).
Each box represents one component of the module
with the assignments shown on the right. 
Arrows represent simple roots of the algebra.
The long diagonal arrows correspond to the middle node
of the Dynkin diagram \protect\figref{fig:dynkin}
while the short horizontal and vertical arrows
correspond to the outer nodes.}
\label{fig:fieldrep}
\end{figure}

A first question is whether the 
symmetry merely constitutes an accidental degeneracy
of the momentum and energy spectrum or whether
it is a symmetry of the full integrable structure. 
Therefore it is useful to look at the eigenvalues of the
commuting charges of the integrable model. The eigenvalues
of the higher local charges 
\[\label{eq:LocalCharges}
Q_r=\frac{1}{r-1}\sum_{j=1}^K
\lrbrk{\frac{i}{(u_j+\ihalf)^{r-1}}-\frac{i}{(u_j-\ihalf)^{r-1}}}
\]
depend on the main Bethe roots $u_j$ only, just like the 
momentum and energy \eqref{eq:MomEng}. 
Consequently their spectrum displays this additional degeneracy.

However, this is not all there is to show;
there are also nonlocal commuting charges whose invariance properties might 
lead to some additional clues. 
Furthermore, the local charge eigenvalues $Q_r$ 
in \eqref{eq:LocalCharges}
are accurate only for $r\leq L$. 
For $r>L$ these charges wrap the spin chain state fully, and they 
receive contributions from the auxiliary Bethe roots $v_j$ and $\dot v_j$. 
This is best seen by considering the transfer matrix in the spin 
representation, which serves as a generating function for the 
local charges as
\[\label{eq:TransGen}
T\indup{spin}(x)=
\exp i\sum_{r=1}^\infty x^{r-1} Q_r.
\]

A transfer matrix is a trace over a particular representation of
the symmetry algebra. 
Therefore, its eigenvalues in a particular representation
are typically written as a sum 
with as many terms as there are 
components in the representation. 
The eigenvalues of a transfer matrix can often be reverse engineered 
by a sort of analytic Bethe ansatz
\cite{Reshetikhin:1983vw}.
This requires some knowledge of the structure of the representations
for which the transfer matrix is to be constructed. 
In particular, it is important to know what the components are
and how they are connected by the simple roots of the algebra.
The structure of the spin representation is depicted in \figref{fig:fieldrep}.
Now it is generally true that the transfer matrix has no dynamic poles,
i.e.~poles whose position depends on the Bethe roots.
Conversely, the terms in the expression for the transfer matrix 
eigenvalue typically have many dynamic poles. These will have 
to cancel between the various terms once the Bethe equations
are imposed. In particular, the Bethe equation for a 
particular type of Bethe root will have to ensure 
the cancellation of singularities between all terms 
that are related by the simple root associated to that Bethe root,
cf.~\figref{fig:dynkin}.
We are then led to the following expression for the transfer matrix
eigenvalue in the spin representation,
see also \cite{Beisert:2005fw,Beisert:2006qh,Belitsky:2007cc},
\<\label{eq:TransSpin}
T\indup{spin}(x)\eq 
\sum_{n=0}^\infty\lrbrk{\frac{x}{x-in}}^L
\prod_{j=1}^M\frac{x-v_j}{x-v_j-in}
\prod_{j=1}^{\dot M}\frac{x-\dot v_j}{x-\dot v_j-in}
\nlnum\nn\qquad\times
\lrbrk{
\delta_{n\neq 0}\prod_{j=1}^K\frac{x-u_j-i(n+\half)}{x-u_j-i(n-\half)}
-2\delta_{n\neq 0}
+\prod_{j=1}^K\frac{x-u_j-i(n-\half)}{x-u_j-i(n+\half)}
}.
\>
We leave it as an exercise for the reader to confirm
the cancellation of poles. 
This is true even if there 
are two coincident auxiliary Bethe roots $v_j=\dot v_{j'}$
in which case a potential double pole is fully eliminated.
Furthermore, it is straightforward to show that the local charge eigenvalues
\eqref{eq:LocalCharges} (for $r \leq L$) follow from 
\eqref{eq:TransGen,eq:TransSpin}
and that only the one term with $n=0$ contributes for $r\leq L$.

This expression is clearly invariant under the degeneracy transformation \eqref{eq:SymBethe}. 
Therefore, the full transfer matrix obeys the enhanced symmetry, which is a 
clear hint that the integrable structure is compatible with the symmetry. 
It is however not fully invariant under it as the eigenvalues of 
transfer matrices in different representations show.
These transfer matrices encode nonlocal charges. 
For instance, for the fundamental and conjugate-fundamental representations
it is easy to construct the transfer matrices
\<
T\indup{fund}(x)\eq 
+\lrbrk{\frac{x+\ihalf}{x}}^L\prod_{j=1}^M\frac{x-v_j-\ihalf}{x-v_j+\ihalf}
\lrbrk{\prod_{j=1}^K\frac{x-u_j+i}{x-u_j}-1}
\nl
+\lrbrk{\frac{x-\ihalf}{x}}^L\prod_{j=1}^{\dot M}\frac{x-\dot v_j+\ihalf}{x-\dot v_j-\ihalf}
\lrbrk{\prod_{j=1}^K\frac{x-u_j-i}{x-u_j}-1}
\>
and
\<
T\indup{\overline{fund}}(x)\eq 
+\lrbrk{\frac{x-\ihalf}{x}}^L\prod_{j=1}^M\frac{x-v_j+\ihalf}{x-v_j-\ihalf}
\lrbrk{\prod_{j=1}^K\frac{x-u_j-i}{x-u_j}-1}
\nl
+\lrbrk{\frac{x+\ihalf}{x}}^L\prod_{j=1}^{\dot M}\frac{x-\dot v_j-\ihalf}{x-\dot v_j+\ihalf}
\lrbrk{\prod_{j=1}^K\frac{x-u_j+i}{x-u_j}-1}.
\>
These expressions are clearly not invariant under the 
shuffling \eqref{eq:SymBethe} of auxiliary Bethe roots.
The violation of the symmetry may be related to the 
fact that the fundamental representations are centrally charged
under $\alg{su}(1,1|2)$
while the spin representation has zero central charge 
and thus belongs to $\alg{psu}(1,1|2)$.%
\footnote{It may be noted that the product 
$T\indup{fund}(x)\,T\indup{\overline{fund}}(x)$
is again invariant under switching the $v$ and $\dot{v}$. 
This is in agreement with the fact that the overall central charge 
for the two representations is zero.}

Finally, we note that the transfer matrix in the spin representation 
\eqref{eq:TransSpin} also has the degeneracy due to the $\alg{psu}(1|1)$ symmetries 
(as do all of the $\mathcal{Q}_r$).  Adding a $v$ or $\dot v$ root 
at zero gives a factor of $x/(x -in)$ in each term of the sum.
This is cancelled by decreasing $L$ by one. 
However, again the degeneracy is not present for the transfer matrix 
in the fundamental or conjugate-fundamental representations.%
\footnote{Their product does not have this degeneracy either.}

\section{Symmetry Enhancement in the Lie Algebra}
\label{sec:LieSym}
\subsection{The Spin Representation}

We begin by describing the spin representation on which the present
spin chain model is based. 
By direct inspection of the explicit expressions we will uncover
an additional $\alg{su}(2)$ symmetry of the model. 

The spin module with Dynkin labels $[0;1;0]$ is spanned by the states,
cf.~\figref{fig:fieldrep}
\[
\state{\phi_a^{(n)}},\qquad
\state{\psi_{\mathfrak{a}}^{(n)}}.
\]
The Latin index $a$ can take values $1,2$,
the Gothic index $\mathfrak{a}$ can
take the values `$\mathord{<}$', `$\mathord{>}$'
and $n$ is a nonnegative integer.
The $\phi$'s are bosonic and the $\psi$'s are fermionic.
In $\superN=4$ gauge theory, these states correspond to the fields
with derivatives
(in the notation of \cite{Beisert:2004ry})
\[\label{eq:StateMap}
\state{\phi_a^{(n)}}\simeq \frac{1}{n!}\,\cder_{11}^n\Phi_{a3},\qquad
\state{\psi_>^{(n)}}\simeq \frac{1}{n!\sqrt{n+1}}\,\cder_{11}^n\Psi_{13},\qquad
\state{\psi_<^{(n)}}\simeq \frac{1}{n! \sqrt{n+1}}\,\cder_{11}^n\dot\Psi^4_1.
\]

The $\alg{psu}(1,1|2)$ algebra has eight supersymmetry generators.
We denote them collectively by $\gen{Q}^{a\beta\mathfrak{c}}$
where a Greek index $\beta$ can take the values `$+$', `$-$'.
In gauge theory the supercharges translate to
\[\label{eq:GenQ}
\begin{array}[b]{rclcrcl}
\gen{Q}^{a+>}=\gen{Q}^a{}_1,&\qquad&
\gen{Q}^{a+<}=\varepsilon^{ab}\dot{\gen{Q}}_{1b},\\[3pt]
\gen{Q}^{a->}=\dot{\gen{S}}^{a1},&\qquad&
\gen{Q}^{a-<}=\varepsilon^{ab}\gen{S}^1{}_b.
\end{array}
\]
At leading order they act on the states as follows, 
\[\label{eq:TransQ}
\begin{array}[b]{rclcrcl}
\gen{Q}^{a+\mathfrak{b}}_{(0)}\state{\phi_c^{(n)}}
\eq \sqrt{n+1}\, \delta^a_c\varepsilon^{\mathfrak{bd}}\state{\psi_{\mathfrak{d}}^{(n)}},
&\quad&
\gen{Q}^{a+\mathfrak{b}}_{(0)}\state{\psi_{\mathfrak{c}}^{(n)}}
\eq\sqrt{n+1}\,\delta^{\mathfrak{b}}_{\mathfrak{c}}\varepsilon^{ad}\state{\phi_{d}^{(n+1)}},
\\[3pt]
\gen{Q}^{a-\mathfrak{b}}_{(0)}\state{\phi_c^{(n)}}
\eq \sqrt{n}\, \delta^a_c\varepsilon^{\mathfrak{bd}}\state{\psi_{\mathfrak{d}}^{(n-1)}},
&\quad&
\gen{Q}^{a-\mathfrak{b}}_{(0)}\state{\psi_{\mathfrak{c}}^{(n)}}
\eq\sqrt{n+1}\,\delta^{\mathfrak{b}}_{\mathfrak{c}}\varepsilon^{ad}\state{\phi_d^{(n)}}.
\end{array}
\]
Furthermore, there are the $\alg{su}(2)$ generators 
$\gen{R}^{ab}=\gen{R}^{ba}$,
which translate to the notation of \cite{Beisert:2004ry} as
$\gen{R}^{ab}=\varepsilon^{ac}\gen{R}^b{}_c$.
They act canonically on the bosonic doublet of states
(to all orders)
\[\label{eq:TransR}
\gen{R}^{ab}\state{\phi_c^{(n)}}
=
\delta^{\{a}_c\varepsilon^{b\}d}\state{\phi_{d}^{(n)}}.
\]
Finally, the $\alg{su}(1,1)$ generators are denoted by 
$\gen{J}^{\alpha\beta}=\gen{J}^{\beta\alpha}$.
They are related to the gauge theory notation as
\[\label{eq:GenJ}
\gen{J}^{++}=\gen{P}_{11},\qquad
\gen{J}^{--}=\gen{K}^{11},\qquad
\gen{J}^{+-}=\half \gen{D}+ \half \gen{L}^1{}_1 + \half \dot{\gen{L}}^1{}_1.
\]
They act on the states by changing the index $n$ by 
up to one unit
\[\label{eq:TransJ}
\begin{array}[b]{rclcrcl}
\gen{J}^{++}_{(0)}\state{\phi_a^{(n)}}
\eq(n+1) \state{\phi_{a}^{(n+1)}},
&\quad&
\gen{J}^{++}_{(0)}\state{\psi_{\mathfrak{a}}^{(n)}}
\eq\sqrt{(n+1)(n+2)} \state{\psi_{\mathfrak{a}}^{(n+1)}},
\\[3pt]
\gen{J}^{+-}_{(0)}\state{\phi_a^{(n)}}
\eq (n + \frac{1}{2}) \state{\phi_{a}^{(n)}},
&\quad&
\gen{J}^{+-}_{(0)}\state{\psi_{\mathfrak{a}}^{(n)}}
\eq (n+1)\state{\psi_{\mathfrak{a}}^{(n)}},
\\[3pt]
\gen{J}^{--}_{(0)}\state{\phi_a^{(n)}}
\eq n \state{\phi_{a}^{(n-1)}},
&\quad&
\gen{J}^{--}_{(0)}\state{\psi_{\mathfrak{a}}^{(n)}}
\eq \sqrt{n(n+1)} \state{\psi_{\mathfrak{a}}^{(n-1)}}.
\end{array}
\]
%

\subsection{The Automorphism}

In the above expressions, the Gothic indices 
$\mathfrak{a},\mathfrak{b},\ldots=\mathord{<},\mathord{>}$
were introduced to handle the two fermionic states in a collective manner. 
The transformation rules \eqref{eq:TransQ,eq:TransR,eq:TransJ}
follow from $\alg{psu}(1,1|2)$ symmetry alone. 
Curiously they can be written with the usual 
index contraction rules using only the 
auxiliary symbols $\delta^{\mathfrak{a}}_{\mathfrak{b}}$
and $\varepsilon^{\mathfrak{ab}}$.
It is therefore obvious that the representation
has an $\alg{su}(2)$ automorphism,
see e.g.~\cite{Gotz:2005ka},
and that the Gothic indices label a doublet of this $\alg{su}(2)$.
We introduce the generators $\gen{B}^{\mathfrak{ab}}$ 
of this $\alg{su}(2),$ which rotate the fermions as
\[\label{eq:TransB}
\gen{B}^{\mathfrak{ab}}\state{\psi_{\mathfrak{c}}^{(n)}}
=\delta^{\{\mathfrak{a}}_{\mathfrak{c}}\varepsilon^{\mathfrak{b}\}\mathfrak{d}}\state{\psi_{\mathfrak{d}}^{(n)}}.
\]

The $\alg{su}(2)$ automorphism can be viewed as an accidental 
symmetry in the $\alg{psu}(1,1|2)$ sector of $\superN=4$ SYM: 
The generators $\gen{B}^{<<}$ and $\gen{B}^{>>}$ transform between
fermions $\Psi$ and conjugate fermions $\dot\Psi$ in gauge theory,
cf.~\eqref{eq:StateMap}.
However, none of the $\alg{psu}(2,2|4)$ generators of the full theory 
acts in such a way.
Only the Cartan generator $\gen{B}^{<>}$ of the $\alg{su}(2)$ automorphism
is equivalent to a combination of the Lorentz generators:
$\gen{B}^{<>}=\gen{L}^{1}{}_{1}-\dot{\gen{L}}^{1}{}_{1}$.

This means we have found an additional symmetry in this sector,
which explains a higher degree of degeneracy in the spectrum. 
Indeed, in terms of the Cartan charges, the transformation of
Bethe roots \eqref{eq:SymBethe} has the same effect as
the generators $\gen{B}^{<<}$ and $\gen{B}^{>>}$.
The two flavors of auxiliary Bethe roots 
$v$ and $\dot v$ effectively form a doublet of the $\alg{su}(2)$ automorphism.%
\footnote{Note, however, that a pair of $v$ and $\dot v$ 
taking the same value form a singlet because of Fermi statistics.}
If there are $M_0$ auxiliary Bethe roots in total, 
the degeneracy is realized as the $M_0$-fold tensor product 
of $\alg{su}(2)$ doublets. 
This tensor product is reducible, and $\alg{su}(2)$ symmetry 
can only account for degeneracy within the irreducible components.
Nevertheless, even the irreducible components turn out to be fully degenerate. 
Therefore, the $\alg{su}(2)$ automorphism explains only part of the extended
degeneracy, and there should be an even larger symmetry. 
This symmetry should have the full tensor product as one irreducible 
multiplet. This behavior is somewhat reminiscent of the Yangian symmetry 
in the Haldane-Shastry model \cite{Haldane:1988gg,SriramShastry:1988gh,Haldane:1992sj},
which also displays fully degenerate tensor products.
We will return to this issue in Section \ref{sec:states},
and consider only the $\alg{su}(2)$ automorphism for the moment.

\subsection{Zero-Momentum States}

As discussed above, 
for zero-momentum states the symmetry is enhanced 
by two copies of $\alg{psu}(1|1)$ with one
central charge.
We shall denote the fermionic generators by
$\hat{\gen{Q}}^{\mathfrak{a}}$ and $\hat{\gen{S}}^{\mathfrak{a}}$
and the central charge by $\hat{\gen{D}}$.
In the gauge theory notation, they represent the supercharges
\[\begin{array}[b]{rclcrcl}
\hat{\gen{Q}}^{<}\eq \dot{\gen{Q}}_{23},
&\qquad&
\hat{\gen{Q}}^{>}\eq -\gen{Q}^4{}_2,
\nln
\hat{\gen{S}}^{<}\eq \gen{S}^2{}_4,
&\qquad&
\hat{\gen{S}}^{>}\eq \dot{\gen{S}}^{32},
\end{array}
\]
and the generator of anomalous dimensions
\[
\hat{\gen{D}}=\half \gen{D} + \gen{L}^2{}_2 + \gen{R}^4{}_4 
= \half \gen{D} + \dot{\gen{L}}^2{}_2 - \gen{R}^3{}_3 =\half \delta \gen{D}.
\]
The last two equalities are satisfied for states within the $\alg{psu}(1,1|2)$ sector. 
The fermionic generators expand in odd powers of the coupling constant,
and they act by increasing or decreasing the length 
of the spin chain by one unit.
At the leading order $\order{g}$, the
generators $\hat{\gen{S}}^{\mathfrak{a}}_{(1)}$
act on two adjacent sites and turn them into a
single site. Explicitly, the action takes the form
\cite{Zwiebel:2005er}
\<\label{TransShat}
\hat{\gen{S}}^{\mathfrak{a}}_{(1)}\state{\phi_{b}^{(m)}\psi_{\mathfrak{c}}^{(n)}}
\eq
-\frac{1}{\sqrt{n+1}}\,
\delta^\mathfrak{a}_\mathfrak{c}\bigstate{\phi_b^{(n+m+1)}},
\nln
\hat{\gen{S}}^{\mathfrak{a}}_{(1)}\state{\psi_{\mathfrak{b}}^{(m)}\phi_{c}^{(n)}}
\eq
\frac{1}{\sqrt{m+1}}\,
\delta^\mathfrak{a}_\mathfrak{b}\bigstate{\phi_c^{(n+m+1)}},
\nln
\hat{\gen{S}}^{\mathfrak{a}}_{(1)}\state{\psi_{\mathfrak{b}}^{(m)}\psi_{\mathfrak{c}}^{(n)}}
\eq
\frac{\sqrt{n+1}}{\sqrt{(m+1)(m+n+2)}}\,
\delta^\mathfrak{a}_\mathfrak{b}\bigstate{\psi_{\mathfrak{c}}^{(n+m+1)}}  
\nl+
\frac{\sqrt{m+1}}{\sqrt{(n+1)(m+n+2)}}\,
\delta^\mathfrak{a}_\mathfrak{c}\bigstate{\psi_{\mathfrak{b}}^{(n+m+1)}}, 
\nln
\hat{\gen{S}}^{\mathfrak{a}}_{(1)}\state{\phi_{b}^{(m)}\phi_{c}^{(n)}}
\eq
\frac{1}{\sqrt{n+m+1}}\,
\varepsilon_{bc} \varepsilon^{\mathfrak{ad}}  \bigstate{\psi_{\mathfrak{d}}^{(n+m)}}.
\>
Conversely, the generators $\hat{\gen{Q}}^{\mathfrak{a}}_{(1)}$
act on a single site and turn it into two,
\<\label{TransQhat}
\hat{\gen{Q}}^{\mathfrak{a}}_{(1)}\state{\phi_{b}^{(n)}}
\eq
\sum_{k=0}^{n-1}
\frac{1}{\sqrt{k+1}}\,
\varepsilon^{\mathfrak{ac}}
\bigstate{\psi_{\mathfrak{c}}^{(k)}\phi_{b}^{(n-1-k)}}
-
\sum_{k=0}^{n-1}
\frac{1}{\sqrt{n-k}}\,
\varepsilon^{\mathfrak{ac}} 
\bigstate{\phi_{b}^{(k)}\psi_{\mathfrak{c}}^{(n-1-k)}},
\nln
\hat{\gen{Q}}^{\mathfrak{a}}_{(1)}\state{\psi_{\mathfrak{b}}^{(n)}}
\eq
\sum_{k=0}^{n-1}
\frac{\sqrt{n-k}}{\sqrt{(k+1)(n+1)}} 
 \, \varepsilon^{\mathfrak{ac}}
\bigstate{\psi_{\mathfrak{c}}^{(k)}\psi_{\mathfrak{b}}^{(n-1-k)}} 
\nl
+
\sum_{k=0}^{n-1}
\frac{\sqrt{k+1}}{\sqrt{(n-k)(n+1)}}  \, \varepsilon^{\mathfrak{ac}}
\bigstate{\psi_{\mathfrak{b}}^{(k)}\psi_{\mathfrak{c}}^{(n-1-k)}} 
\nl
-
\sum_{k=0}^n
\frac{1}{\sqrt{n+1}}\,
\delta^\mathfrak{a}_\mathfrak{b}  \varepsilon^{cd}
\bigstate{\phi_{c}^{(k)}\phi_{d}^{(n-k)}}.
\>
Again, by inspection the representations of $\alg{psu}(1|1)^2$ 
turns out to have a manifest $\alg{su}(2)$ automorphism.
It is nice to see that the unified treatment of the two fermionic states as a doublet
compresses the expressions found in \cite{Zwiebel:2005er} somewhat. 
Furthermore, when the construction of \cite{Zwiebel:2005er}
is to be carried to higher perturbative orders 
one may expect the $\alg{su}(2)$ symmetry to reduce the 
number of permitted terms and thus simplify the analysis. 
Finally, we should note that there is a unique lift of 
the action \eqref{TransShat,TransQhat} to the nonplanar level.
This means that the nonplanar $\alg{psu}(1,1|2)$ sector
of $\superN=4$ SYM will also have the additional $\alg{su}(2)$ symmetry.%
\footnote{It is likely, however, that the $2^{M_0}$ degeneracy
will be lifted into the irreducible components of $\alg{su}(2)$.}

\subsection{The Hamiltonian}

In the zero-momentum sector, the Hamiltonian $\ham = \delta \gen{D}= 2 \, \hat{\gen{D}}$ 
is given by twice the anticommutator of $\alg{psu}(1|1)^2$ supercharges, see \eqref{eq:psu11comm}.%
\footnote{Because the extended $\alg{psu}(1|1)^2$ central charges vanish on zero-momentum states,
it does not matter which pair of conjugate supercharges we use to compute $\hat{\gen{D}}$.} 
For a length $L$ zero-momentum state, 
this can be written purely in terms of two-site to two-site interactions as follows
\< \label{eq:twositeham}
\ham \eq \sum_{j=1}^{L} \ham(j,j+1),
\nln
\ham(j,j+1) \eq 2 \, \hat{\gen{Q}}^<(j) \hat{\gen{S}}^>(j, j+1)
+ 2 \, \hat{\gen{S}}^>(j-1, j) \hat{\gen{Q}}^<(j) + 2 \, \hat{\gen{S}}^>(j+1, j+2) \hat{\gen{Q}}^<(j)
\nl
+ \hat{\gen{S}}^>(j, j+1) \hat{\gen{Q}}^<(j)+ \hat{\gen{S}}^>(j+1, j+2) \hat{\gen{Q}}^<(j+1).
\>
The arguments of the supercharges refer to the sites of the spin chain on which the supercharges act. 
The generator $\hat{\gen{S}}^{\mathfrak{a}}(j, j+1)$ replaces 
the fields at sites $j$ and $j+1$ with a new (sum of) field(s) 
at site $j$, and $\hat{\gen{Q}}^{\mathfrak{b}}(j)$ 
acts in the conjugate way. 
From the last equality, 
one can compute the explicit interactions of $\ham(j,j+1)$. 
Then \eqref{eq:twositeham} also gives $\ham$ for 
periodic states with \emph{arbitrary} momentum, 
as this definition does not require cyclic states. 
This Hamiltonian for general periodic states 
still commutes with the (leading order) $\alg{psu}(1,1|2)$ generators, 
is integrable, and for a given Bethe eigenstate has eigenvalue equal
to the energy $E$ determined by the Bethe equations \eqref{eq:Bethe} and by \eqref{eq:MomEng}.

Using $\gen{R}$ and $\gen{B}$ symmetry, 
as well as the fact that the Hamiltonian has even parity, 
these interactions can be written in terms of seven functions.%
\footnote{Parity, or $\mathbf{p}$, reverses the order of the fields 
in spin chain states. In addition to minus signs for every resulting crossing 
of fermionic fields, parity also includes a factor of $(-1)^L$ 
for states of length $L$. All of the extended $\alg{psu}(1,1|2)$ and $\alg{psu}(1|1)^2$ 
generators have $\mathbf{p}$ eigenvalue $+1$, or are parity even.} 
We now give the explicit form of $\ham(1,2)$ in a hermitian basis%
\footnote{Alternatively, one could eliminate square roots by using 
a different normalization for the fermionic fields, 
at the expense of no longer having a hermitian basis. 
In that case one appearance of $f_3$ would be replaced with a new eighth function.}
 \<
  \ham \state{\phi_a^{(j)} \phi_b^{(n-j)}} \eq
\sum_{k=0}^{n}  f_1 (n, j,k) \state{\phi_a^{(k)} \phi_b^{(n-k)}}
 +  \sum_{k=0}^{n} f_2 (n, j,k)\state{\phi_b^{(k)} \phi_a^{(n-k)}} 
\nl
+ \sum_{k=0}^{n-1} f_3 (n, j,k)\varepsilon_{ab} \varepsilon^{\mathfrak{cd}} \state{\psi_{\mathfrak{c}}^{(k)} \psi_{\mathfrak{d}}^{(n-1-k)}},
  \nln
  \ham \state{\phi_a^{(j)} \psi_{\mathfrak{b}}^{(n-j)}} \eq
\sum_{k=0}^{n} f_4 (n, j,k) \state{\phi_a^{(k)} \psi_{\mathfrak{b}}^{(n-k)}} 
+\sum_{k=0}^{n}  f_5 (n, j,k)\state{\psi_{\mathfrak{b}}^{(k)} \phi_a^{(n-k)}},
  \nln
   \ham \state{\psi_{\mathfrak{a}}^{(j)} \phi_b^{(n-j)}} \eq
\sum_{k=0}^{n} f_4 (n,n-j,n-k) \state{\psi_{\mathfrak{a}}^{(k)} \phi_b^{(n-k)}}
 +\sum_{k=0}^{n}  f_5 (n,n-j,n-k)\state{\phi_b^{(k)} \psi_{\mathfrak{a}}^{(n-k)}},
  \nln
 \ham \state{\psi_{\mathfrak{a}}^{(j)} \psi_{\mathfrak{b}}^{(n-j)}} \eq
\sum_{k=0}^{n} f_6 (n, j,k) \state{\psi_{\mathfrak{a}}^{(k)} \psi_{\mathfrak{b}}^{(n-k)}}
 +
\sum_{k=0}^{n}  f_7 (n, j,k)\state{\psi_{\mathfrak{b}}^{(k)} \psi_{\mathfrak{a}}^{(n-k)}}
\nl
+ \sum_{k=0}^{n+1} f_3 (n+1,k,j)\varepsilon_{\mathfrak{ab}} \varepsilon^{cd} \state{\phi_c^{(k)} \phi_d^{(n+1-k)}},
  \>
with the coefficient functions $f_n$
  \<
  f_1(n, j, k) \eq 2 \, \delta_{jk}\bigbrk{h(j)+h(n-j)}-\frac{2 \, \delta_{j\neq k}}{|j-k|}+\frac{2}{n+1}\,,
  \nln
  f_2(n, j, k) \eq -\frac{2}{n+1}\,,
  \nln
  f_3(n, j, k) \eq \frac{2\,(n-k)}{(n+1)\sqrt{k+1}\sqrt{n-k}}-\frac{2\,\theta(j-k-1)}{\sqrt{k+1}\sqrt{n-k}}\,,
  \nln
  f_4(n, j, k) \eq 2 \, \delta_{jk}\bigbrk{h(j)+h(n-j+1)}+2 \, \theta(k-j-1)\frac{\sqrt{n-k+1}}{(j-k)\sqrt{n-j+1}}
  \nl
          -2 \, \theta(j-k-1)\frac{\sqrt{n-j+1}}{(j-k)\sqrt{n-k+1}}\,,
  \nln
  f_5(n, j, k) \eq -\frac{2 \, \theta(k-j)}{\sqrt{n-j+1}\sqrt{k+1}}\,,
 \nln[20pt]
  f_6(n, j, k) \eq 2 \, \delta_{jk}\left(h(j+1)+h(k+1)-\frac{1}{n+2}\right)
  \nl
  + 2 \, \frac{\sqrt{j+1}\sqrt{n-j+1}}{\sqrt{k+1}\sqrt{n-k+1}} \left( \theta(k-j-1)\frac{(n-j+k+2)(n-k+1)}{(n-j+1)(j-k)(n+2)}  \right.
  \nl
  \left.  \qquad\qquad\qquad\qquad-\theta(j-k-1) \frac{(k+1)(n+j-k+2)}{(j+1)(j-k)(n+2)}\right),  
  \\\nn
  f_7(n, j, k) \eq 2 \, \frac{\sqrt{j+1}\sqrt{n-j+1}}{\sqrt{k+1}\sqrt{n-k+1}} \left(\frac{n-k+1}{(n-j+1)(n+2)}-\frac{\theta(j-k-1)(j-k)}{(j+1)(n-j+1)} \right). 
 \>
The symbol $\theta(n)$ represents the step function, 
which is one for $n \geq 0$ and 0 otherwise,
and $h(n)$ is the $n$-th harmonic number,
  \[
  h(n) = \sum_{j=1}^n \frac{1}{j}\,.
  \]
  %

\section{Some Degenerate States}
\label{sec:states}

Let us now consider the full observed degeneracy.
We will try to get acquainted with it by constructing explicitly some
degenerate states. 
Here and in the following sections
we will work only at leading order in the coupling constant $g$. 
In other words, the $\alg{psu}(1,1|2)$ generators 
$\gen{Q},\gen{J}$
are truncated at $\order{g^0}$, and for the $\alg{psu}(1|1)^2$
generators $\hat{\gen{Q}},\hat{\gen{S}}$ we take only the $\order{g^1}$
contributions $\hat{\gen{Q}}_{(1)},\hat{\gen{S}}_{(1)}$
in \eqref{TransShat,TransQhat}.

\subsection{Vacuum} \label{sec:vac}

The simplest state that is
part of a nontrivial multiplet is 
\[\label{eq:vac}
\state{0_L}=
\state{\psi^{(0)}_<\psi^{(0)}_<\psi^{(0)}_<\ldots \psi^{(0)}_<}.
\]
We shall call it the vacuum state of length $L$. 
Note that it is not the ground state of the model, 
but it is a homogeneous eigenstate of
the Hamiltonian, and we can place excitations on it by
flipping some of the spins.
In the above Bethe ansatz it is represented by 
$K=L$ main Bethe roots and
$M=L$ auxiliary Bethe roots.
The roots are the solutions to the algebraic equations
(including $u=\infty$ and twice $v=\infty$)
\[
(u+\ihalf)^L=(u-\ihalf)^L,\qquad
(v+i)^L+(v-i)^L=2v^L.
\]
The equation for the main Bethe roots can be solved explicitly as
$u_k=\half\cot(\pi k/L)$.
The momentum and energy of this state are given by
\eqref{eq:MomEng}
\[P=\pi (L-1),\qquad E=4L.\]
The eigenvalue of the transfer matrix in the spin representation reads
\eqref{eq:TransSpin}
\[
T\indup{spin}(x)=1+
\sum_{n=0}^\infty\frac{x^L\bigbrk{2x^L-(x+i)^L-(x-i)^L}}{(x-i n)^L(x-in-i)^L}\,.
\]

Note that for even $L$ the overall momentum is maximal, 
$P\equiv \pi$, while for odd $L$ the overall momentum is zero,
$P\equiv 0$.
Therefore only the states with odd $L$ are physical states 
of AdS/CFT, and only for those the symmetry algebra enlarges by 
$\alg{psu}(1|1)^2$.

The vacuum state is part of a $\alg{su}(2)$ multiplet of $L+1$ \emph{states}. 
The $L+1$ components are given by $(\gen{B}^{<<})^{0,1,\ldots,L}\state{0_L}$.
Note also that it is part of a multiplet of $L-1$ \emph{multiplets} of $\alg{psu}(1,1|2)$.%
\footnote{Due to the $\alg{su}(2)$ grading of the $\alg{psu}(1,1|2)$ 
algebra these two numbers differ by two.}
The $L-1$ highest-weight components are obtained by acting 
with the cubic operator given in \appref{sec:cubic};
they read $((\gen{J}^3)^{<<})^{0,1,\ldots,L-2}\state{0_L}$.

\subsection{Degenerate Eigenstates}

Let us now consider the set of states where the flavor of one auxiliary Bethe root
is flipped. One can convince oneself that a state 
is composed from basis states of the typical form
\[
\gen{Q}^{2-<}(k)\,
\gen{Q}^{1-<}(l)\,
\gen{J}^{++}(m)\,
\state{0_L}\sim
\state{\ldots\MMM{\phi}{k}{}^{(0)}_1\ldots\MMM{\phi}{l}{}^{(0)}_2\ldots 
\MMM{\psi}{m}{}^{(1)}_<\ldots}. \label{typicalstate}
\]
The arguments of the generators correspond to the sites 
of the spin chain on which they should act.
Here we have only displayed the excitations 
while the vacuum sites $\psi^{(0)}_<$ have been suppressed.
The operators $\gen{J}(k)$ act as
the leading order generators in \eqref{eq:TransQ,eq:TransJ}
on site $k$ of the chain.%
\footnote{The statistics of the fermionic
generators $\gen{Q}(k)$ is taken into account 
by first permuting it to its place of action. This
may cause a sign flip.}
Note that if two or all of the three excitations coincide on a single site
they will give rise to $\phi^{(1)}_1$, $\phi^{(1)}_2$ or $\psi^{(0)}_>$.
We find precisely $L+1$ states of this form 
completely degenerate with the vacuum $\state{0_L}$.
Three of these states are descendants of $\alg{psu}(1,1|2)$,
\[
\varepsilon_{ab}\gen{Q}^{a-<}\gen{Q}^{b-<}\gen{J}^{++}\state{0_L},\quad
\varepsilon_{ab}\gen{Q}^{a-<}\gen{Q}^{b+<}\state{0_L},\quad
\varepsilon_{ab}\gen{Q}^{a+<}\gen{Q}^{b-<}\state{0_L},
\]
and one is the $\alg{su}(2)$ descendant
\[
\gen{B}^{<<}\state{0_L}.
\]
However, since $\gen{B}^{\mathfrak{ab}}$  does not commute 
with $\alg{psu}(1,1|2)$, it is more convenient to use instead 
the cubic operator $(\gen{J}^3)^{\mathfrak{ab}}$ presented 
in \appref{sec:cubic} (built from cubic combinations of ordinary $\alg{psu}(1,1|2)$ 
and $\alg{su}(2)$ generators),
\[
(\gen{J}^3)^{\mathfrak{<<}}\state{0_L}.
\]
The generator $(\gen{J}^3)^{\mathfrak{ab}}$ commutes
with $\alg{psu}(1,1|2)$ and 
therefore moves between  $\alg{psu}(1,1|2)$ highest weight states.

For even $L$ (and nonzero momentum) this exhausts the set of 
trivial descendants. There remain $L-3$ unexplained degenerate states.
For odd $L$ the vacuum is a zero-momentum state, and therefore the 
additional $\alg{psu}(1|1)^2$ symmetry applies. It yields one further
descendant,
\[
\hat{\gen{S}}^<\hat{\gen{Q}}^<\state{0_L}. \label{odddescendant}
\]
Consequently there are only $L-4$ unexplained degenerate states
in this case.

Among the remaining degenerate states we find one state 
with the very simple form
\<\label{eq:simplestate}
\state{1_L}\eq
\sum_{n,k=1}^{L}(-1)^{k}
\gen{J}^{++}(k+n)\,
\varepsilon_{ab}
\gen{Q}^{a-<}(1+n)\,
\gen{Q}^{b-<}(L+n)\,\state{0_L}
\nl
- (1 - (-1)^L)\gen{B}^{<<}\, \state{0_L}.
\>
One can confirm straightforwardly that it is a highest-weight state 
of $\alg{psu}(1,1|2)$.
For even length this state is indeed linearly independent of
the above descendants.
For odd length, however, the state is proportional to the
$\alg{psu}(1|1)^2$ descendant \eqref{odddescendant},
$\state{1_L}\sim \hat{\gen{S}}^<\hat{\gen{Q}}^<\state{0_L}$.
This turns out to be a special case because of the overall momentum being zero. 
We will return to this issue in the next section.

We have also found a second degenerate state with a slightly more complicated form,
\<
\label{eq:nextsimplest}
\state{2_L}\eq 
\sum_{n,k=1}^{L}(-1)^{k} (2 k - L -1 + \delta_{k1} - \delta_{kL})
\gen{J}^{++}(k+n)\,
\varepsilon_{ab}
\gen{Q}^{a-<}(1+n)\,
\gen{Q}^{b-<}(L+n)\,
\state{0_L} 
\nl
+\sum_{k=2}^{L} \sum_{n=1}^{L}(-1)^{k}
\gen{J}^{++}(k+n)\,
\varepsilon_{ab}
\gen{Q}^{a-<}(2+n)\,
\gen{Q}^{b-<}(L+n)\,
\state{0_L} 
\nl
+
(1 + (-1)^L)(L-1)\gen{B}^{<<}\, \state{0_L}.
\>
This state is also a highest weight state of $\alg{psu}(1,1|2)$, 
and for odd length is not a $\alg{psu}(1|1)^2$ descendant
of $\state{0_L}$.

\subsection{Parity} \label{sec:parity}

The degenerate states do not all have the same parity.
For $L$ even or odd we find $\half (L-2)$ or $\half (L-3)$ states, respectively,
which have opposite parity than the vacuum.%
\footnote{The definition of parity may also
include shifts $\shift^k$ of the chain
which act nontrivially on states with overall momentum. 
It is therefore more convenient to only
specify the parity w.r.t.~a reference state.} 
Recalling the above results, this means that after removing 
the trivial descendants 
there is always one more degenerate state with opposite
parity than with equal parity.
More explicitly, we can say that 
$\state{1_L}$ has the opposite parity as 
$\state{0_L}$ for even $L$ and the same
parity as $\state{0_L}$ for odd $L$.
Conversely, the state $\state{2_L}$ has the same parity as 
$\state{0_L}$ for even $L$ and the opposite
parity as $\state{0_L}$ for odd $L$.

\section{Nonlocal Symmetry} 
\label{Non-local}

To account for the additional degeneracy, 
it is natural to seek new symmetry generators.
We will take into account the findings regarding the Bethe ansatz and 
the form of the degenerate states found 
in the previous section to construct some nonlocal generators $\yang$.
We will then investigate their algebra.

\subsection{Bilocal Generators}

First of all, an elementary step between two degenerate Bethe states
consists in changing the flavor of one auxiliary Bethe root,
as discussed in Section \ref{sec:BetheSym}. 
The $\alg{su}(2)$ generators $\gen{B}^{\mathfrak{ab}}$
qualitatively act in the same way. 
This indicates that the new generators 
will be in the same representation, 
i.e.~in the adjoint/spin-one/triplet representation of $\alg{su}(2)$.
We will thus denote them by $\yang^{\mathfrak{ab}}=\yang^{\mathfrak{ba}}$.

As the example degenerate states given in the previous section 
have multiple nonadjacent excitations,
we should look for nonlocal generators.
The simplest degenerate state $\state{1_L}$ 
in \eqref{eq:simplestate}
has a pair of adjacent excitations and a single
excitation that is not near the pair. 
A generator that creates such a state from the vacuum $\state{0_L}$ 
consequently has to be bilocal (at least). 
More complicated states with multilocal excitations
such as $\state{2_L}$ in \eqref{eq:nextsimplest} could in principle 
be generated by repeated application of these bilocal generators.

Furthermore, we know that the form of the example degenerate state $\state{1_L}$ 
in \eqref{eq:simplestate}
is qualitatively identical to the second order $\alg{psu}(1|1)^2$ descendant 
$\hat{\gen{S}}^{\mathfrak{<}}\hat{\gen{Q}}^{\mathfrak{<}}\state{0_L}$.
Thus we expect $\yang^{\mathfrak{ab}}$ to act similarly
to $\hat{\gen{S}}^\mathfrak{\{a}\hat{\gen{Q}}^\mathfrak{b\}}$.

Here we have to make a distinction between states 
with zero and states with nonzero momentum.
For zero momentum the combination 
$\hat{\gen{S}}^\mathfrak{\{a}\hat{\gen{Q}}^\mathfrak{b\}}$
already explains the degenerate state $\state{1_L}$. 
However, due to the $\alg{psu}(1|1)^2$ algebra, it cannot
explain any of the other degenerate states.
Conversely, in the case of nonzero momentum the individual generators 
$\hat{\gen{S}}^\mathfrak{a}$ and $\hat{\gen{Q}}^\mathfrak{b}$ 
cannot be defined independently because it is not possible to change 
the length of the spin chain preserving the momentum.%
\footnote{The eigenvalues of a lattice momentum operator 
take the values $2\pi m/L$ (mod $2\pi$).
Changing the length $L$ by one unit only preserves
the eigenvalue zero.}
It is nevertheless possible to consistently define the
product $\hat{\gen{S}}^\mathfrak{\{a}\hat{\gen{Q}}^\mathfrak{b\}}$ 
for nonzero-momentum states because it preserves the length.
This is the bilocal operator
\[
\yang^\mathfrak{ab} =  \sum_{j=0}^{L-1} \sum_{i=0}^{L+1} 
(1 - \half\delta_{i,0} - \half\delta_{i,L+1}) 
\,\shift^{j-i} \, \hat{\gen{S}}^{\{\mathfrak{a}}(1,2)\,  
\shift^i \, \hat{\gen{Q}}^{\mathfrak{b}\}} (1) \, \shift^{-j} . 
\label{eq:nonlocalgen1}  
\]
Here, $\shift$ is the operator that shifts the chain by one site to the right;
it commutes with all of the local symmetry generators. 
The summation over $j$ ensures that $\yang^\mathfrak{ab}$ acts homogeneously on the chain, 
and the symmetrization in the indices makes it a $\alg{su}(2)$ triplet, 
as needed to explain the degeneracy. 
The generator $\hat{\gen{Q}}(1)$ removes the first site of the chain and replaces 
it with two sites, while $\hat{\gen{S}}(1,2)$ 
replaces the first two sites of the spin chain with one. 
So, the generator $\yang^\mathfrak{ab}$ consists of products of the $\hat{\gen{Q}}$ and $\hat{\gen{S}}$ 
generators acting all possible distances apart, 
with equal weight except for a symmetric regularization 
when a $\hat{\gen{S}}$ interaction acts on both sites created by a $\hat{\gen{Q}}$ interaction. 
The regularization resolves the one-site ambiguity in where to place newly created sites.

For zero-momentum states 
the action of $\yang^{\mathfrak{ab}}$
is equivalent to the action of $\hat{\gen{S}}^{\{\mathfrak{a}}\hat{\gen{Q}}^{\mathfrak{b} \}}$. 
Therefore, it cannot be used to immediately explain the additional degeneracy 
beyond the established $\alg{psu}(1|1)^2$ symmetry
in the zero-momentum sector. 
We will discuss this further in Section \ref{sec:double}.
However, $\yang^{\mathfrak{ab}}$
does commute exactly with $\alg{psu}(1,1|2)$ and with the Hamiltonian
even if the momentum is nonzero;
a proof is given in Appendix \ref{app:biproof}.
Therefore the existence of $\yang^{\mathfrak{ab}}$ proves the
additional degeneracies for all states with
nonzero momentum. 

The generators $\yang^{\mathfrak{ab}}$ 
immediately explain the form of the simplest
degenerate state \eqref{eq:simplestate} found in the last 
section; it is related to the vacuum by applying 
$\yang^{<<}$ once, 
\[
\state{1_L}\sim\yang^{<<}\state{0_L}.
\]
For even length $L\leq 10$ we have checked directly that the 
remaining descendants are given by
\[\label{eq:furtherstates}
\yang^{<>}\state{1_L}, \quad\ldots,\quad  (\yang^{<>})^{(L-4)}\state{1_L}.
\]
Of course, further application of $\yang^{<>}$ generates no additional 
linearly independent states. Since the $(\yang^{<>})^{m}\state{1_L}$ 
include all of the degenerate states, there is a linear combination of them 
that equals the degenerate state generated by the cubic invariant $(\gen{J}^3)^{<<}\state{0_L}$. 
Also, a short computation shows that the $\yang^{<>}$ transform under parity $\mathbf{p}$ as
\[
\mathbf{p} \, \yang^{\mathfrak{ab}} \, \mathbf{p} = \shift \yang^{\mathfrak{ab}} .
\]
This is consistent with the counting of the parities of degenerate states 
done in Section \ref{sec:parity}. 
When acting on the even-length vacuum ($\shift$ eigenvalue $-1$), 
the $\yang^{\mathfrak{ab}}$ are parity odd 
and generate a sequence of alternating parity degenerate states.

For the odd-length states, which have vanishing momentum,
one can easily convince oneself 
using $\yang^{ab}\simeq\hat{\gen{S}}^{\{\mathfrak{a}} \hat{\gen{Q}}^{\mathfrak{b}\}}$
that the states 
\eqref{eq:furtherstates} are all proportional to $\state{1_L}$.

\subsection{An Infinite-Dimensional Algebra}
\label{sec:loop}

Let us first understand the algebra of $\yang^{\mathfrak{ab}}$
in the zero-momentum sector, where we have a
representation in terms of $\alg{psu}(1|1)^2$ generators.
It is not difficult to convince oneself of the following relations,
\<
\bigcomm{\gen{B}^{\mathfrak{ab}}}
        {(-\hat{\gen{D}})^m \hat{\gen{S}}^{\mathfrak{c}}\hat{\gen{Q}}^{\mathfrak{d}}}
\eq
(-\hat{\gen{D}})^m \varepsilon^{\mathfrak{c\{b}}\hat{\gen{S}}^{\mathfrak{a\}}}\hat{\gen{Q}}^{\mathfrak{d}}
-(-\hat{\gen{D}})^m \hat{\gen{S}}^{\mathfrak{c}}\hat{\gen{Q}}^{\mathfrak{\{b}}\varepsilon^{\mathfrak{a\}d}},
\nln
\bigcomm{(-\hat{\gen{D}})^m \hat{\gen{S}}^{\mathfrak{a}}\hat{\gen{Q}}^{\mathfrak{b}}}
        {(-\hat{\gen{D}})^n \hat{\gen{S}}^{\mathfrak{c}}\hat{\gen{Q}}^{\mathfrak{d}}}
\eq
(-\hat{\gen{D}})^{m+n+1} 
\varepsilon^{\mathfrak{cb}}\hat{\gen{S}}^{\mathfrak{a}}\hat{\gen{Q}}^{\mathfrak{d}}
-(-\hat{\gen{D}})^{m+n+1} 
\varepsilon^{\mathfrak{ad}}\hat{\gen{S}}^{\mathfrak{c}}\hat{\gen{Q}}^{\mathfrak{b}}.
\>
Denoting these combinations by 
$\yang^{\mathfrak{ab}}_k$, $k=0,1,2,\ldots$,
such that
$\yang^{\mathfrak{ab}}_0=\gen{B}^{\mathfrak{ab}}$ and
$\yang^{\mathfrak{ab}}_n\simeq(-\hat{\gen{D}})^{n-1}\yang^{\mathfrak{ab}}$,
we obtain the infinite-dimensional algebra
\[\label{eq:yalg}
\comm{\yang_m^{\mathfrak{ab}}}{\yang_n^{\mathfrak{cd}}}
=
\varepsilon^{\mathfrak{cb}}\yang_{m+n}^{\mathfrak{ad}}
-\varepsilon^{\mathfrak{ad}}\yang_{m+n}^{\mathfrak{cb}}
.
\]
This algebra is a parabolic subalgebra
of the loop algebra of $\alg{su}(2)$.

We conjecture that the same algebra \eqref{eq:yalg} holds 
not only for the zero-momentum sector, but for all states
if we identify 
\[
\yang^{\mathfrak{ab}}_0=\gen{B}^{\mathfrak{ab}},\qquad
\yang^{\mathfrak{ab}}_1=\yang^{\mathfrak{ab}},\qquad
\yang^{\mathfrak{ab}}_{n+1}=
-\half\varepsilon_{\mathfrak{cd}}\comm{\yang^{\mathfrak{c\{a}}}{\yang_{n}^{\mathfrak{b\}d}}}.
\]
It is quite clear that the relations with $m=0$ or $n=0$ hold by $\alg{su}(2)$ symmetry.
Furthermore, the relation with $m=n=1$ merely defines $\yang^{\mathfrak{ab}}_{2}$. 
The relations with $m+n\geq 3$ are nontrivial and have to be verified. 

In fact, the relations with $m+n=3$ are the Serre relations for the algebra,
and they imply all the relations with $m+n>3$. 
In the following we will prove this statement by induction. 
For convenience, we switch to an adjoint basis for $\yang_n^{\mathfrak{i}}$,
$\mathfrak{i}=1,2,3$ where the $\alg{su}(2)$ structure constants
are given by the totally antisymmetric tensor $\varepsilon^{\mathfrak{ijk}}$.
The commutation relations can now be written for all nonnegative integer levels $N$ as
\[
\label{eq:yangalgebra}
\comm{\yang_m^{\mathfrak{i}}}{\yang_{N-m}^{\mathfrak{j}}} =
 \varepsilon^{\mathfrak{ijk}} \yang_{N}^{\mathfrak{k}} , \quad m=0,\ldots N. 
\]
Assume \eqref{eq:yangalgebra} is satisfied at some level $N\geq 3$. 
Then we use five main steps to show that it is satisfied at level $N+1$.

\begin{bulletlist}
\item
Step 1. Using our inductive assumption, consider the equations 
for $m=1,\ldots N-2$ and their cyclic permutations,
\<
0\eq \comm{\yang_1^2}{\yang_{N-m}^1}+  \comm{\yang_1^1}{\yang_{N-m}^2} 
\nln\eq \bigcomm{\yang_m^3}{\comm{\yang_1^2}{\yang_{N-m}^1}} + 
 \bigcomm{\yang_m^3}{ \comm{\yang_1^1}{\yang_{N-m}^2}} 
\nln\eq \comm{\yang_1^2}{\yang_N^2}-\comm{\yang_{m+1}^1}{\yang_{N-m}^1}
+\comm{\yang_{m+1}^2}{\yang_{N-m}^2}-\comm{\yang_1^1}{\yang_N^1} .
\>
Comparing the $m=M$ and $m=N-M-1$ equations, we find that 
\[
\comm{\yang_m^1}{\yang_{N+1-m}^1}=\comm{\yang_m^2}{\yang_{N+1-m}^2}
=\comm{\yang_m^3}{\yang_{N+1-m}^3}, \quad m=1,\ldots N.
\]

\item
Step 2. We also have, for $m=1,\ldots N$
\<
\comm{\yang_m^1}{\yang_{N+1-m}^1} \eq \bigcomm{\yang_m^1}{\comm{\yang_1^2}{\yang_{N-m}^3}} 
\nln\eq 
\comm{\yang_{m+1}^3}{\yang_{N-m}^3}-\comm{\yang_1^2}{\yang_N^2},
\>
and cyclic permutations. Using the result from step 1, we find
\[
\comm{\yang_m^1}{\yang_{N+1-m}^1}= m \comm{\yang_1^1}{\yang_{N}^1}, \quad m=1, \ldots N,
\]
and similarly for cyclic permutations. However, since 
\[
\comm{\yang_1^1}{\yang_{N}^1}= - \comm{\yang_N^1}{\yang_{1}^1}.
\]
we must have
\[
0=\comm{\yang_m^1}{\yang_{N+1-m}^1}= \comm{\yang_m^2}{\yang_{N+1-m}^2}
= \comm{\yang_m^3}{\yang_{N+1-m}^3}, \quad m=1, \ldots N.
\]

\item
Step 3. Commuting $\yang_0$ with $\comm{\yang_m^1}{\yang_{N+1-m}^1}$ 
(and cyclic permutations) yields
\[
\comm{\yang_m^{\mathfrak{i}}}{\yang_{N+1-m}^{\mathfrak{j}}} 
= - \comm{\yang_m^{\mathfrak{j}}}{\yang_{N+1-m}^{\mathfrak{i}}}, \quad m=1,\ldots N.
\]

\item
Step 4. We can now show that there is a unique consistent way 
to define $\yang_{N+1}^{\mathfrak{k}}$. 
For instance, consider the following equations for $m=1,\ldots N-1$,
\<
\comm{\yang_m^1}{\yang_{N+1-m}^2}
\eq\bigcomm{\yang_m^1}{\comm{\yang_1^3}{\yang_{N-m}^1}} 
= -\comm{\yang_{m+1}^2}{\yang_{N-m}^1} 
\nln
\eq \comm{\yang_{m+1}^1}{\yang_{N-m}^2}
\nln
\earel{\ldots}
\nln
\eq \comm{\yang_{N}^1}{\yang_{1}^2}
\nln
\eq \yang_{N+1}^3.
\>

\item
Step 5. It is now straightforward to use any of the 
equivalent expressions for $\yang_{N+1}^{\mathfrak{k}}$ to check that
\[
\comm{\yang_0^\mathfrak{i}}{\yang_{N+1}^\mathfrak{j}}
=\varepsilon^{\mathfrak{ijk}}\yang_{N+1}^{\mathfrak{k}}.
\]
This completes the set of equations at level $N+1$. 
Therefore, assuming the level-3 equations are satisfied, 
\eqref{eq:yangalgebra} is satisfied for all $N$. 
\end{bulletlist}

At this time, a direct proof of the level-3 relations 
is beyond our technical capabilities. 
Note that to prove the level-3 relations, 
it is sufficient to check that (switching back to the previous $\alg{su}(2)$ notation)
$\comm{\yang_1^{<>}}{\yang_2^{<>}}=0$,
since commutators with the $\gen{B}$ yield the remaining relations.
This relation can also be written using only bilocal generators as
\[
\bigcomm{\yang^{<>}}{\comm{\yang^{<<}}{\yang^{>>}}}=0.
\]
Still, we have to gain confidence in the level-3 relations.
As a start, using \texttt{Mathematica} we have checked 
that they are satisfied on many states of small excitation number, 
including all states of length 4 with 4 or fewer excitations 
(above the half-BPS vacuum) and all state of length 5 or 6 with 3 or fewer excitations. 
Also checked were states with larger lengths and excitation numbers, 
including a length-7, 7-excitation state. 
Checking much longer or higher excitation states rapidly becomes impractical
because of combinatorics. 
However, we consider the evidence described above as persuasive. 
Hopefully, a complete proof will become possible in the future.

\subsection{The Representation of the Loop Algebra}
\label{sec:rep}

The observed degeneracies motivating this work should correspond 
to irreducible $2^M$-dimensional representations of the above loop algebra.
Finite-dimensional representations of loop algebras are 
typically tensor products of evaluation representations.
In an evaluation representation, the level-$n$ generator $\yang_n$
acts like the level-0 generator $\yang_0$ multiplied by the 
$n$-th power of the evaluation parameter $x$
\[
\yang_n \state{x} = x^n \yang_0 \state{x}.
\]
Tensor products of evaluation representations $\state{x_k}$
with distinct evaluation parameters $x_k$
are generally irreducible. The basic reason is that
the sum over $(x_k)^n$ is not proportional
to the $n$-th power of the sum over $x_k$.

In our case the relevant evaluation module is 
two-dimensional and consists of the states
\[
\state{<,x} \quad \text{and} \quad \state{>, x}.
\]
Explicitly, the generators act on these states as 
(note that $\yang_0^{\mathfrak{ab}}=\gen{B}^{\mathfrak{ab}}$)
\[\label{eq:evalrep}
\begin{array}[b]{rclrcl}
  \yang_n^{<<} \state{<,x} \eq + x^n \state{>,x},  
& \yang_n^{<<} \state{>,x} \eq   0, \\[3pt]
  \yang_n^{>>} \state{<,x} \eq 0, 
& \yang_n^{>>} \state{>,x} \eq -x^n \state{<,x} ,  \\[3pt]
  \yang_n^{<>} \state{<,x} \eq -\half x^n \state{<,x}, 
& \yang_n^{<>} \state{>,x} \eq +\half x^n \state{>,x},
\end{array}
\]
which is consistent with the algebra \eqref{eq:yalg}.
Then, tensor products labeled by the highest-weight state
\[
\state{\Psi}=
\state{<,x_1}\otimes\state{<,x_2}\otimes\ldots\otimes\state{<,x_{M}}
\]
with distinct $x_k$
form multiplets of dimension $2^M$. 
These multiplets are characterized by the eigenvalues of the generator $\yang_n^{<>}$
\[ 
\yang_n^{<>} \state{\Psi}
=-\frac{1}{2} \left( \sum_{i=1}^M x_k^n \right)  
\state{\Psi}.
\]
\begin{table}\centering
$ \begin{array}[tbp]{|c |c| c| c| c|}\hline
L & p  & u & v & \dot{v} \\ \hline \hline
3 & \pm \frac{2 \pi}{3}  &1 \mp 4 \sqrt{3} u - 12 u^2 \pm 16 \sqrt{3} u^3 & 1 \pm \sqrt{3} v - 6 v^2 & 1 \mp \sqrt{3} \dot{v}   \\ \hline
4 & \pm \frac{\pi}{2}  & 1 \mp 16 u - 40 u^2 \pm  64 u^3 + 80 u^4 & 1 \pm 5 v - 6 v^2 \mp 10 v^3 & 1\pm \dot{v}
 \\  \hline
4 & \pi  & u - 4 u^3 & 1 - 6 v^2  & \text{---}
 \\  \hline
5 & \pm \frac{2 \pi} {5} & -\half \sqrt{1 + \frac{2}{\sqrt{5}}}(1- 40 u^2 + 80 u^4) & 1 - 15 v^2 + 15 v^4   &   \sqrt{1 + \frac{2}{\sqrt{5}}} \mp \dot{v}
 \vspace{-.1cm} \\
 &  &\pm 3 u \mp 40 x^3 \pm 48 u^5  & \pm 5\sqrt{1 + \frac{2}{\sqrt{5}}} (v - 2 v^3)  & \\ \hline
6 & \pi  & -3 u + 40 u^3 - 48 u^5  &  1 - 15 v^2 + 15 v^4 & \text{---}
 \\ \hline
8 & \pi  & u - 28 u^3 + 112 u^5 - 64 ^7 & 1 - 28 v^2 + 70 v^4 - 28 v^6 & \text{---} 
\\\hline
\end{array}$
\caption{Eigenstates used for checking the relationship \protect\eqref{eq:relyangbethe} 
between $\yang$ eigenvalues and Bethe roots. 
The first two columns give the length and momentum of the eigenstates. 
The last three columns give the polynomials whose zeros are the Bethe roots. 
Note that the contributions of equal auxiliary roots ($v_k=\dot{v}_j$) 
cancel in the expression for $\yang_n^{<>}$ \protect\eqref{eq:yangneigenvalue}.} 
\label{Table:eigenstates}
\end{table} 
In fact, by examining some representative eigenstates listed in 
Table \ref{Table:eigenstates}, we find that the $x_k$ 
should be simply related to the auxiliary Bethe roots $v_k$ and $\dot v_k$ as
\[ \label{eq:relyangbethe}
x_k = \frac{i (1 - e^{iP})}{v_k}\,,
\]
where $P$ is the overall momentum of the state.
With this identification, the algebra implies that any nonzero momentum Bethe 
eigenstate $\state{\Psi}$ characterized by auxiliary roots $\{v_1, \ldots ,v_M\}$ 
and $\{\dot{v}_1, \ldots ,\dot{v}_{\dot{M}}\}$ satisfies 
\[ \label{eq:yangneigenvalue}
\yang_n^{<>} \state{\Psi} = -\half \bigbrk{i (1 - e^{iP})}^n\left( \sum_{k=1}^M \frac{1}{v_k^n} 
- \sum_{k=1}^{\dot{M}} \frac{1}{\dot{v}_k^n} \right) \state{\Psi}.
\]
This identification \eqref{eq:relyangbethe} is not surprising since 
the auxiliary Bethe roots are closely associated with the degeneracy. 
Furthermore, the inverse dependence on the $v$ and on the $\dot{v}$ 
follows from \eqref{eq:evalrep}.
This is necessary for compatibility with invariance of $\yang^{\mathfrak{ab}}$ 
under the $\alg{psu}(1,1|2)$ algebra.
It is also consistent with the fact that a pair of equal 
auxiliary Bethe roots $v$ and $\dot v$
leads to a singlet rather than a quadruplet. 

It is curious that the overall momentum $P$ appears in the 
definition of the evaluation parameter.
It actually cancels the singularities 
that occur when there are auxiliary roots $v$ or $\dot{v}$ at zero: 
As explained in \secref{sec:BetheSymText} 
this can only happen for zero-momentum states,
and in that case the factor in the numerator $(1 - e^{iP})$ also goes to zero. 
The explicit evaluation of $\yang^{<>}_n$ for the odd-length
vacuum state $\state{0_L}$ in \eqref{eq:vac} gives the proper regularization (for $n > 0$)
\footnote{The result can in fact be derived from a regularization of \eqref{eq:yangneigenvalue} as well:
Assume $v_1=0$ and $e^{iP}=1$ for some solution to the Bethe equations.
Take a small deformation of the set of Bethe roots 
which preserves the Bethe equation for $v_1$.
Then the limit (as $P$ returns toward 0) of the combination $i(1-e^{iP})/v_1$ equals $-\half E$,
which equals the eigenvalue of $-\hat{\gen{D}}$.
We thank the referee for pointing out this method to us.}
\[
\yang^{<>}_n\state{0_L}
=-\half (-\hat D)^n\state{0_L}.
\]
Here $\hat D=\half E=2L$ is the eigenvalue of $\hat{\gen{D}}$
which equals half the energy of the state.
In other words, the state corresponds to the following tensor product of 
evaluation representations,
\[
\state{0_L}=
\state{<,- \hat D}
\otimes
\state{<,0}
\otimes
\ldots
\otimes
\state{<,0}.
\]
Therefore the generators $\yang_n$, $n>0$, transform effectively only the 
first doublet. This is fully consistent with our above findings
that the generators $\yang_n$ cannot explain the degeneracy in the
zero-momentum case and also with 
the algebra $\yang_n\simeq (-\hat{\gen{D}})^{n-1}\yang$.
  
Finally, we should emphasize that we have not proven 
that the identification \eqref{eq:relyangbethe} is satisfied for all states,
but we have given compelling evidence of its truth.

\subsection{Relation to Yangian Symmetry}
 \label{sec:yangian}

The apparent asymptotic integrability of the $\superN=4$ SYM spin chain is equivalent
to the existence of a Yangian symmetry \cite{Drinfeld:1985rx}, 
which is a nonlocal infinite-dimensional symmetry.
The Yangian of the $\superN=4$ SYM spin chain was constructed 
at leading order in \cite{Dolan:2003uh,Dolan:2004ps}, 
and its perturbative corrections in subsectors have been studied 
in \cite{Serban:2004jf,Agarwal:2004sz,Agarwal:2005ed,Zwiebel:2006cb,Beisert:2006xx}.
The Yangian is a Hopf algebra whose structure is 
the subject of many recent investigations
\cite{Gomez:2006va,Plefka:2006ze,Beisert:2006qh,Gomez:2007zr,
Torrielli:2007mc,Young:2007wd,Beisert:2007ds,Moriyama:2007jt}.
In general, for a Lie algebra with generators $\gen{J}^A$, the Yangian 
is generated by the Lie generators $\gen{J}^A_0=\gen{J}^A$ 
combined with additional generators, $\gen{J}^A_1$.  
In a spin chain description, the $\gen{J}^A_1$ act as bilocal products 
of the $\gen{J}^A_0$
\[ \label{eq:coproduct}
\gen{J}^A_1\simeq
\sum_{k<n}
f^A{}_{BC} \gen{J}_0^B(k) \gen{J}_0^C(n),
\]
where $f^A{}_{BC}$ are the structure constants. From this
action, it is clear that the $\gen{J}^A_1$ transform 
in the adjoint of the Lie algebra. The Yangian generators must satisfy a 
Serre relation,%
\footnote{The symmetric triple product is $\{ x_1, x_2, x_3 ] = \sum_{i \neq j \neq k} x_i x_j x_k$, 
with appropriate additional signs for fermionic $x$.}
\< 
\biggcomm{\gen{J}_1^{A}}{\gcomm{\gen{J}_1^{B}}{\gen{J}_0^{C}}} 
- \biggcomm{\gen{J}_0^{A}}{\gcomm{\gen{J}_1^{B}}{\gen{J}_1^{C}}}  \eq 
  \sfrac{1}{6} a^{ABC}{}_{DEF} \, \{ \gen{J}_0^D, \gen{J}_0^E, \gen{J}_0^F ], \nln
a^{ABC}{}_{DEF} \eq (-1)^{(EM)} f^{AK}{}_{D} f^B{}_{E}{}^{L} f^C{}_{F}{}^{M}  f_{KLM}
 \label{eq:serre} .
\>
The term on the right hand side implies that a Yangian is a 
deformation of the loop (sub)algebra of a Lie algebra. Also, combining this Serre relation 
with the adjoint transformation of the $\tilde{\gen{J}}^A$ implies 
another relation,
\<\label{eq:serre2}
\biggcomm{\gcomm{\gen{J}_1^{A}}{\gen{J}_1^B}}{\gcomm{\gen{J}_0^P}{\gen{J}_1^{Q}}}
+ \biggcomm{\gcomm{\gen{J}_1^{P}}{\gen{J}_1^Q}}{\gcomm{\gen{J}_0^A}{\gen{J}_1^{B}}} 
 \eq  \sfrac{1}{6} a^{ABC}{}_{DEF} f^{PQ}{}_C \{ \gen{J}_0^D, \gen{J}_0^E, \gen{J}_1^F ]  
.
\>
This second Serre relation is useful when considering a $\alg{su}(2)$ algebra, 
since in that case the first Serre relation is trivial.

Let us now compare the action of $\yang^{\mathfrak{ab}}$ in \eqref{eq:nonlocalgen1}
to the formal action of Yangian generators in \eqref{eq:coproduct}.
The former acts as a bilocal product of  
$\hat{\gen{S}}^{\mathfrak{a}}$ and $\hat{\gen{Q}}^{\mathfrak{b}}$.
The same is true for the $\alg{su}(2)$ automorphism
$\hat{\gen{B}}^{\mathfrak{ab}}$ of $\alg{psu}(1|1)^2$
with vanishing central charges $\hat{\gen{C}}^{\mathfrak{ab}}=0$.
In fact, the action is equal up to  
(not yet investigated) issues related to the length-changing nature of the
$\alg{psu}(1|1)^2$ generators
$\hat{\gen{S}}^{\mathfrak{a}}$ and $\hat{\gen{Q}}^{\mathfrak{b}}$.
Therefore it is natural to identify our generators
$\yang^{\mathfrak{ab}}$ with the level-one generators
of the $\alg{su}(2)$ automorphism%
\footnote{As in \cite{Zwiebel:2006cb}, the definition of the bilocal product 
here needs to be generalized naturally to allow for multisite 
(and even length-changing) symmetry generators. 
With appropriate modifications of the local terms, 
we could write the bilocal product also including 
terms with $\hat{\gen{S}}$ acting first, in agreement with the bilocal action.}
\[
\yang^{\mathfrak{ab}}=\hat{\gen{B}}_1^{\mathfrak{ab}}.
\]
The generators $\yang^{\mathfrak{ab}}$ would thus
enlarge the $\alg{psu}(1|1)^2$ Yangian 
(which is a part of the full $\alg{psu}(2,2|4)$ Yangian \cite{Dolan:2003uh,Dolan:2004ps})
by an automorphism
in just the same way as the generators
$\gen{B}^{\mathfrak{ab}}$ enhance the 
the $\alg{psu}(1|1)^2$ Lie algebra by the $\alg{su}(2)$ automorphism.
Consistently with this identification, the Serre relation \eqref{eq:serre2} 
implies that the $\yang^{\mathfrak{ab}}$ generate 
an \emph{undeformed} $\alg{su}(2)$ loop (sub)algebra, 
since the relevant combinations of structure constants appearing 
on the right side vanishes for central charges 
$\gen{C}^{\mathfrak{ab}}=0$.%
\footnote{This Yangian relation thus provides an efficient way to 
see that the level-3 relation suffices to guarantee that the loop 
subalgebra is satisfied. 
The level-3 relation is equivalent to the Serre relation 
for the $\alg{su}(2)$ part of the Yangian.}

Of course, for nonzero momentum states the $\alg{psu}(1|1)^2$ symmetry no longer applies. 
However, it does apply for infinite-length states 
(which are typically required also for Yangian symmetry to be realized%
\footnote{The bilocal Yangian generators usually cannot be defined 
consistently with periodic boundary conditions.}%
). 
We can then view the loop algebra symmetry for nonzero-momentum states 
simply as the consequence of the extended $\alg{psu}(1|1)^2$ Yangian 
for the infinite-length chain combined with the fact that
the $\yang=\hat{\gen{B}}_1$ are length-preserving. 
In contrast, the other Yangian generators, 
the $\hat{\gen{Q}}_1$ or $\hat{\gen{S}}_1$, 
clearly are not a symmetry for finite-length nonzero-momentum states 
since they also change the length of the chain. 
It is still unusual even for part of this Yangian symmetry 
to be realized exactly by the Hamiltonian for finite-length states.  
It is closely tied to the fact that the generators
$\hat{\gen{S}}$ and $\hat{\gen{Q}}$ change the length by a definite and opposite amount,
so that a bilocal product consistent with periodic boundary conditions can be constructed. 

While we identify the $\alg{su}(2)$ automorphisms 
of the $\alg{psu}(1,1|2)$ and $\alg{psu}(1|1)^2$ algebras, 
$\gen{B}=\hat{\gen{B}}$,
apparently the corresponding Yangians cannot be identified,
$\gen{B}_1\neq\hat{\gen{B}}_1$.
For instance, the $\yang=\hat{\gen{B}}_1$ commute with the $\alg{psu}(1,1|2)$ algebra, 
while the Yangian generator $\gen{B}_1$ should not commute. 
However, it would still be interesting to generalize \cite{Zwiebel:2006cb} 
to obtain the $\mathcal{O}(g^2)$ corrections to the extended $\alg{psu}(1,1|2)$ Yangian. 
It is possible still that there is some relation between the $\mathcal{O}(g^2)$ Yangian generators 
of the two automorphisms. 

Finally, note that we can now expect contributions to other Yangian generators 
at $\mathcal{O}(g^2)$ that have similar spin chain structure as the $\yang^{\mathfrak{ab}}$. 
Suitable sectors that have generators  that act nontrivially at $\mathcal{O}(g)$ 
include the $\alg{su}(2|3)$ sector as well as the full $\alg{psu}(2,2|4)$ spin chain.

\subsection{Zero-Momentum Degeneracy and the Yangian Double}
\label{sec:double}

With our new understanding of the origin of the algebra generated by the $\yang$, 
there is a natural explanation for the remaining degeneracy of the zero-momentum sector. 
As noted above, any Bethe eigenstate in the zero-momentum sector has a root at $v=0$ or $\dot{v}=0$. 
The contribution from this root dominates the eigenvalue of the $\yang^{<>}$, 
so that these states only form doublets of the loop algebra. 
What is needed to explain the other degenerate states is a generator with the inverse eigenvalues, 
for which the nonzero roots would dominate. 
This is precisely what we would expect from the full $\alg{su}(2)$ loop algebra, 
which would follow from the double Yangian \cite{Drinfeld:1986in} 
for the extended $\alg{psu}(1|1)^2$ algebra. 

The full $\alg{su}(2)$ loop algebra takes the same form as in \eqref{eq:yalg}, 
except now the $\yang_m$ are defined for all integer values $m$. 
So the full algebra is generated by 
$\yang_0^{\mathfrak{ab}}$, $\yang_1^{\mathfrak{ab}}$, and $\yang_{-1}^{\mathfrak{ab}}$. 
The additional relations that would need to be checked to verify 
that the $\yang_{-1}^{\mathfrak{ab}}$ generate the rest of the algebra
include the level-$(-3)$ Serre relation and 
\[
\comm{\yang_{-1}^{\mathfrak{ab}}}{\yang_{1}^{\mathfrak{cd}}} = \varepsilon^{\mathfrak{cb}}\gen{B}^{\mathfrak{ad}}
-\varepsilon^{\mathfrak{ad}}\gen{B}^{\mathfrak{cb}}.
\]
It would be very interesting to find the spin chain representation 
for the $\yang_{-1}^{\mathfrak{ab}}$, 
which basically invert the $\yang_{1}^{\mathfrak{ab}}$. 
We leave this investigation for the future, 
but note that using the example states in Table \ref{Table:eigenstates} 
and \eqref{eq:yangneigenvalue} with $n=-1$ will 
provide significant information about these generators. 
However, unlike the $\yang_1^{\mathfrak{ab}}$ there does not appear
to be as natural a representation in terms of ordinary symmetry generators.
Finally, once one finds the $\yang_{-1}^{\mathfrak{ab}}$, 
one could immediately compute the $\hat{\gen{Q}}_{-1}$ 
and $\hat{\gen{S}}_{-1}$, which would not 
act just as products of ordinary symmetry generators.

\subsection{A Singlet Bilocal Generator}

It is curious to note that there exists a  bilocal
generator $\yangsing$ very similar to the $\yang^{\mathfrak{ab}}$, which is a $\alg{su}(2)$-singlet
\[\label{eq:nonlocalgen2}  
\yangsing =  \sum_{j=0}^{L-1} \sum_{i=0}^{L+1} (1 - \half\delta_{i,0} - \half\delta_{i,L+1}) 
\half\varepsilon_{\mathfrak{ba}}
\,\shift^{j-i} \, \hat{\gen{S}}^{\mathfrak{a}}(1,2)\,  \shift^i \, \hat{\gen{Q}}^{\mathfrak{b}} (1) \, \shift^{-j} . 
\]
Like the $\yang^{\mathfrak{ab}}$, it commutes with the $\alg{psu}(1,1|2)$ algebra and the one-loop Hamiltonian,
as discussed in Appendix \ref{app:biproof}.

Similar to the reasoning used at the beginning of Section \ref{sec:loop},
we can use the zero-momentum reduction 
for $\yangsing$ to conjecture that 
it commutes with the $\yang^{\mathfrak{ab}}_n$ for all $n$.
Again, we have obtained very strong evidence using \texttt{Mathematica}. 
We have checked that these commutators vanish for the same set of states 
described in the last paragraph of Section \ref{sec:loop}. 
It is however presently not clear how to generalize 
$\yangsing$ to an infinite-dimensional algebra of $\yangsing_m$. 
For such an algebra, we would have $\yangsing_0=\hat{\gen{A}}$, 
and for zero-momentum states the remaining generators 
would simply be $\yangsing_n=\hat{\gen{D}}^{n-1}\yangsing$, 
yielding an abelian algebra that commutes with the $\yang_n^{\mathfrak{ab}}$.

Using the states in Table \ref{Table:eigenstates}, 
we find that the eigenvalues of $\yangsing$ only differ from those of $\yang_1^{<>}$ 
by the relative sign between the $v$ and $\dot{v}$ contributions, 
\[
\yangsing \state{\Psi} = -\ihalf (1 - e^{iP})\left( \sum_{k=1}^M \frac{1}{v_k} 
+ \sum_{k=1}^{\dot{M}} \frac{1}{\dot{v}_k} \right) \state{\Psi}.
\]
From this and the fact that $\yangsing$ reduces to a product of $\alg{psu}(1|1)^2$ 
symmetry generators for zero-momentum states, we see that $\yangsing$ 
does not map between different $\alg{psu}(1,1|2)$ multiplets.

Similar arguments as in Section \ref{sec:yangian} imply that we can identify $\yangsing$ 
as a bilocal Yangian generator, $\yangsing=\hat{\gen{A}}_1$, 
and the Serre relation \eqref{eq:serre}
then implies that $\yangsing$ commutes with the triplet $\yang$. 
Finally, $\yangsing$ also should have a double, 
but the double would commute with $\yangsing$.

\section{Conclusions and Outlook} \label{sec:conc}

In this article we have investigated a curious $2^M$-fold degeneracy
of an integrable spin chain with $\alg{psu}(1,1|2)$ symmetry.
This degeneracy was observed at the level of Bethe equations
in \cite{Beisert:2005fw}. Here we have considered the symmetry algebra
that explains the degeneracy. We have constructed
two triplets of symmetry generators, $\gen{B}$ and $\yang$, 
at the level of operators acting on spin chain states. 
The local generators $\gen{B}$ form a $\alg{su}(2)$
automorphism of $\alg{psu}(1,1|2)$ while the 
bilocal generators $\yang$ commute with $\alg{psu}(1,1|2)$.
Together they apparently generate a
subalgebra of the loop algebra of $\alg{su}(2)$.
This extended symmetry algebra commutes with the Hamiltonian 
and thus explains the degeneracy.

It remains an open problem to identify the spin chain operators 
that generate the $2^M$ 
degeneracy for zero-momentum states. As argued above,
 these operators are likely to generate the remaining part 
of the full $\alg{su}(2)$ loop algebra. It is possible that 
these operators' spin chain representation is not simple. 
However, this still deserves further study especially 
because it is also possible that they would give new insight 
into the origin of the simple next-to-leading order corrections 
to the local symmetry generators obtained in \cite{Zwiebel:2005er}.

While we have restricted our study to the one-loop Hamiltonian, 
it is clear that the symmetry enhancement persists at higher loops. 
The $2^M$ degeneracy of the Bethe ansatz is preserved 
by the higher-loop corrections \cite{Beisert:2005fw}.  
Therefore, we expect the $\yang^{\mathfrak{ab}}$ to receive loop 
corrections so that they commute with the loop-corrected Hamiltonian. 
Note that the leading terms for the bilocal symmetry generators 
$\yang_{(2)}^{\mathfrak{ab}}$
discussed in this paper correspond to $\order{g^2}$. 
Given the Yangian origin of the $\yang^{\mathfrak{ab}}$,
we expect the corrections for the bilocal generators to involve 
substituting the appropriate loop corrections 
for the $\alg{psu}(1|1)^2$ generators appearing in the expression 
for the $\yang^{\mathfrak{ab}}$, similar to the quantum corrections 
to bilocal Yangian generators studied in \cite{Zwiebel:2006cb}.
That is, at $\order{g^{2\ell}}$ the bilocal generators should take the form
\[
\yang_{(2\ell)}^{\mathfrak{ab}} \simeq
\sum_{m=1}^{\ell} \sum_{j=0}^{L-1} \sum_{i=0}^{L+1} 
\,\shift^{j-i} \, \hat{\gen{S}}_{(2m-1)}^{\{\mathfrak{a}}(1,\ldots, m+1)\,  
\shift^i \, \hat{\gen{Q}}^{\mathfrak{b}\}}_{(2\ell-2m+1)} (1,\ldots,\ell-m+1) \, \shift^{-j} .
\]
Explicit calculation will be required to find the regularization of the overlap 
between $\hat{\gen{S}}$ and $\hat{\gen{Q}}$. 
The study of these corrections may be very useful 
in constraining the higher-loop contributions to the local symmetry generators.

As at leading order, for cyclic states the $\yang_{(2\ell)}^{\mathfrak{ab}}$ 
will reduce to ordinary products of the (loop-corrected) $\alg{psu}(1|1)^2$ generators.  
Therefore, we expect that the loop corrections 
will also preserve the algebra of the $\yang_n^{\mathfrak{ab}}$.  
This would be consistent with the loop algebra following from 
the extended $\alg{psu}(1|1)^2$ Yangian of the infinite-length chain, 
as discussed in Section \ref{sec:yangian}, 
which is expected also to all orders in perturbation theory.

The degeneracy was observed in the context of 
$AdS_5\times S^5$ string theory. However, it might also
be relevant for certain superstring models on 
$AdS_3\times S^3$ or $AdS_2\times S^2$ which also possess
$\alg{psu}(1,1|2)$ symmetry. Further suitable models
include the principal chiral/WZW model on the group manifold 
$\widetilde{\grp{PSU}}(1,1|2)$ or some of it cosets. 
For instance, in some of these cases an additional $\grp{su}(2)$
and some even larger unexplained degeneracies 
were noticed in \cite{Gotz:2006qp}. 
It is conceivable that they are of a similar origin as the ones discussed here.

\subsection*{Acknowledgments}

We would like to thank A.~Kleinschmidt and V.~Schomerus for interesting discussions.
B.~Z.~would like to thank the Albert Einstein Institute Potsdam for 
hospitality during the course of this work.  
The work of N.~B.~was supported in part by the 
U.S.~National Science Foundation Grant No.~PHY02-43680
and by the Alfred P.~Sloan Foundation. 
The work of B.~Z.~has been supported by a 
National Science Foundation Graduate Research Fellowship. 
Any opinions, findings and conclusions or recommendations expressed in
this material are those of the authors and 
do not necessarily reflect the views of the National Science Foundation.

\appendix

\section{Commutation Relations}
\label{sec:algebra}

In the following we shall list the commutation relations for the symmetry algebras. 

\subsection{Maximally Extended $\alg{psu}(1,1|2)$ Algebra}

Let us first consider the 
$\alg{psu}(1,1|2)$ algebra. 
It consists of the three $\alg{su}(2)$ generators 
$\gen{R}^{ab}=\gen{R}^{ba}$,
the three $\alg{su}(1,1)$ generators 
$\gen{J}^{\alpha\beta}=\gen{J}^{\beta\alpha}$,
and the eight fermionic generators
$\gen{Q}^{a\beta\mathfrak{c}}$.
All Latin, Greek and Gothic indices
can take one out of two values.
A summary of commutation relations reads
\<
\comm{\gen{R}^{ab}}{\gen{R}^{cd}}\eq 
\varepsilon^{cb}\gen{R}^{ad}
-\varepsilon^{ad}\gen{R}^{cb},
\nln
\comm{\gen{J}^{\alpha\beta}}{\gen{J}^{\gamma\delta}}\eq 
\varepsilon^{\gamma\beta}\gen{J}^{\alpha\delta}
-\varepsilon^{\alpha\delta}\gen{J}^{\gamma\beta},
\nln
\comm{\gen{R}^{ab}}{\gen{Q}^{c\delta\mathfrak{e}}}\eq 
\half \varepsilon^{ca}\gen{Q}^{b\delta\mathfrak{e}}
+\half \varepsilon^{cb}\gen{Q}^{a\delta\mathfrak{e}},
\nln
\comm{\gen{J}^{\alpha\beta}}{\gen{Q}^{c\delta\mathfrak{e}}}\eq 
\half \varepsilon^{\delta\alpha}\gen{Q}^{c\beta\mathfrak{e}}
+\half \varepsilon^{\delta\beta}\gen{Q}^{c\alpha\mathfrak{e}},
\nln
\acomm{\gen{Q}^{a\beta\mathfrak{c}}}{\gen{Q}^{d\epsilon\mathfrak{f}}}
\eq 
\varepsilon^{\beta\epsilon}\varepsilon^{\mathfrak{cf}}
\gen{R}^{da}
-\varepsilon^{ad}\varepsilon^{\mathfrak{cf}}
\gen{J}^{\beta\epsilon}
+\varepsilon^{ad}\varepsilon^{\beta\epsilon}
\gen{C}^{\mathfrak{c}\mathfrak{f}}.
\>
For completeness, we have introduced 
a maximal set of three central charges 
$\gen{C}^{\mathfrak{ab}}=\gen{C}^{\mathfrak{ba}}$.
In the case of the spin representation they act trivially.
The algebra furthermore admits an $\alg{su}(2)$ grading. 
The commutators with the generators $\gen{B}^{\mathfrak{ab}}=\gen{B}^{\mathfrak{ba}}$
of the automorphism are canonical,
\<
\comm{\gen{B}^{\mathfrak{ab}}}{\gen{B}^{\mathfrak{cd}}}\eq 
\varepsilon^{\mathfrak{cb}}\gen{B}^{\mathfrak{ad}}
-\varepsilon^{\mathfrak{ad}}\gen{B}^{\mathfrak{cb}},
\nln
\comm{\gen{B}^{\mathfrak{ab}}}{\gen{Q}^{c\delta\mathfrak{e}}}\eq 
\half \varepsilon^{\mathfrak{ea}}\gen{Q}^{c\delta\mathfrak{b}}
+\half \varepsilon^{\mathfrak{eb}}\gen{Q}^{c\delta\mathfrak{a}},
\nln
\comm{\gen{B}^{\mathfrak{ab}}}{\gen{C}^{\mathfrak{cd}}}\eq 
\varepsilon^{\mathfrak{cb}}\gen{C}^{\mathfrak{ad}}
-\varepsilon^{\mathfrak{ad}}\gen{C}^{\mathfrak{cb}}.
\>
Note that the ``central charges'' $\gen{C}^{\mathfrak{ab}}$ 
now become a spin-1 triplet under this $\alg{su}(2)$ automorphism,
i.e.~they are not central for the maximally extended algebra.
All in all this algebra can be denoted as $\alg{su}(2)\ltimes\alg{psu}(1,1|2)\ltimes \Reals^3$.

\subsection{Maximally Extended $\alg{psu}(1|1)^2$ Algebra}

The only nontrivial commutator of the $\alg{psu}(1|1)^2$ algebra reads
\< \label{eq:psu11comm}
\acomm{\hat{\gen{Q}}^{\mathfrak{a}}}{\hat{\gen{S}}^{\mathfrak{b}}}
\eq 
\hat{\gen{C}}^{\mathfrak{a}\mathfrak{b}}
+\varepsilon^{ab}\hat{\gen{D}}.
\>
For completeness we have introduced a triplet $\hat{\gen{C}}^{\mathfrak{ab}}$ 
of central charges to accompany the singlet $\hat{\gen{D}}$. 
In our spin chain model the triplet acts trivially, 
$\hat{\gen{C}}^{\mathfrak{ab}}=0$. 

The algebra admits a $\alg{u}(2)$ grading, 
which can be split up into $\alg{su}(2)$ 
and $\alg{u}(1)$ gradings. 
The $\alg{su}(2)$ automorphism 
is defined by the commutation relations
\<
\comm{\hat{\gen{B}}^{\mathfrak{ab}}}{\hat{\gen{B}}^{\mathfrak{cd}}}\eq 
\varepsilon^{\mathfrak{cb}}\hat{\gen{B}}^{\mathfrak{ad}}
-\varepsilon^{\mathfrak{ad}}\hat{\gen{B}}^{\mathfrak{cb}},
\nln
\comm{\hat{\gen{B}}^{\mathfrak{ab}}}{\hat{\gen{Q}}^{\mathfrak{c}}}\eq 
\half \varepsilon^{\mathfrak{ca}}\hat{\gen{Q}}^{\mathfrak{b}}
+\half \varepsilon^{\mathfrak{cb}}\hat{\gen{Q}}^{\mathfrak{a}},
\nln
\comm{\hat{\gen{B}}^{\mathfrak{ab}}}{\hat{\gen{S}}^{\mathfrak{c}}}\eq 
\half \varepsilon^{\mathfrak{ca}}\hat{\gen{S}}^{\mathfrak{b}}
+\half \varepsilon^{\mathfrak{cb}}\hat{\gen{S}}^{\mathfrak{a}},
\nln
\comm{\hat{\gen{B}}^{\mathfrak{ab}}}{\hat{\gen{C}}^{\mathfrak{cd}}}\eq 
\varepsilon^{\mathfrak{cb}}\hat{\gen{C}}^{\mathfrak{ad}}
-\varepsilon^{\mathfrak{ad}}\hat{\gen{C}}^{\mathfrak{cb}},
\>
while the $\alg{u}(1)$ grading $\hat{\gen{A}}$
distinguishes $\hat{\gen{Q}}$ from $\hat{\gen{S}}$,
\<
\comm{\hat{\gen{A}}}{\hat{\gen{Q}}^{\mathfrak{a}}}\eq 
+\hat{\gen{Q}}^{\mathfrak{a}},
\nln
\comm{\hat{\gen{A}}}{\hat{\gen{S}}^{\mathfrak{a}}}\eq 
-\hat{\gen{S}}^{\mathfrak{a}}.
\>
Altogether the algebra can be denoted by $\alg{u}(2)\ltimes \alg{psu}(1|1)^2\ltimes \Reals^4$.

A priori the $\alg{su}(2)$ automorphisms $\gen{B}$ and $\hat{\gen{B}}$ of 
$\alg{psu}(1,1|2)$ and $\alg{psu}(1|1)^2$, respectively, 
are not identical, 
but they merely satisfy the same commutation relations. 
For the spin representation of the product of these algebras 
in perturbative gauge theory, they should however be identified
$\gen{B}=\hat{\gen{B}}$.

The $\alg{psu}(1|1)^2$ algebra can be embedded 
in another $\alg{psu}(1,1|2)$ algebra, 
with the fermionic generators now written as 
$\hat{\gen{Q}}^{a \beta \mathfrak{c}}$.
Then we have
\[\begin{array}[b]{rclcrcl}
\hat{\gen{Q}}^{\mathfrak{a}} \eq \hat{\gen{Q}}^{1 + \mathfrak{a}},
&\quad&
\hat{\gen{A}}\eq \hat{\gen{R}}^{12},
\\[3pt]
\hat{\gen{S}}^{\mathfrak{a}} \eq \hat{\gen{Q}}^{2 - \mathfrak{a}},
&\quad&
\hat{\gen{D}}\eq -\hat{\gen{J}}^{+-}+\hat{\gen{R}}^{12}.
\end{array}
\]

\section{Multilinear Operators} \label{app:multilinear}

In this appendix we list some relevant multilinear operators for
the symmetry algebra. These include the quadratic Casimir invariant, 
but also an interesting triplet of cubic operators. 
We then show that the cubic operators satisfy the same 
algebra as the $\yang^{\mathfrak{ab}}$ and can be used 
to deform the $\yang^{\mathfrak{ab}}$ while preserving 
this algebra.

\subsection{Quadratic Invariants}
\label{sec:casimir}

It is straightforward to construct the quadratic Casimir 
for the maximally extended $\alg{psu}(1,1|2)$ algebra
introduced in Appendix \ref{sec:algebra},
\[\label{eq:Casimir112}
\gen{J}^2=
2\varepsilon_{\mathfrak{bc}}
 \varepsilon_{\mathfrak{da}}
\gen{B}^{\mathfrak{ab}}\gen{C}^{\mathfrak{cd}}
+\varepsilon_{bc}\varepsilon_{da}
 \gen{R}^{ab}\gen{R}^{cd}
-\varepsilon_{\beta\gamma}\varepsilon_{\delta\alpha}
 \gen{J}^{\alpha\beta}\gen{J}^{\gamma\delta}
-\varepsilon_{ad}\varepsilon_{\beta\epsilon}\varepsilon_{\mathfrak{cf}}
 \gen{Q}^{a\beta\mathfrak{c}}\gen{Q}^{d\epsilon\mathfrak{f}}.
\]
For the algebra without central extensions, $\gen{C}^{\mathfrak{ab}}=0$,
the first terms simply drops out. 
The centrally extended algebra without automorphism, on the other hand,
does not have a quadratic invariant because the first term 
is important, but it requires $\gen{B}^{\mathfrak{ab}}$.

For the maximally extended $\alg{psu}(1|1)^2$ the quadratic Casimir 
operator reads
\[\label{eq:Casimir11}
\hat{\gen{J}}^2=
\varepsilon_{\mathfrak{bc}}\varepsilon_{\mathfrak{da}}
\acomm{\hat{\gen{B}}^{\mathfrak{ab}}}{\hat{\gen{C}}^{\mathfrak{cd}}}
+
\acomm{\hat{\gen{A}}}{\hat{\gen{D}}}
-\varepsilon_{\mathfrak{ab}}
\comm{\hat{\gen{Q}}^{\mathfrak{a}}}{\hat{\gen{S}}^{\mathfrak{b}}}.
\]

In the combined algebra of $\alg{psu}(1,1|2)$ and $\alg{psu}(1|1)^2$
with identified automorphisms 
$\gen{B}^{\mathfrak{ab}}=\hat{\gen{B}}^{\mathfrak{ab}}$
also the central charges have to be identified,
$\gen{C}^{\mathfrak{ab}}=\hat{\gen{C}}^{\mathfrak{ab}}$,
in order for a quadratic invariant to exist. 
This invariant is the sum of 
\eqref{eq:Casimir112}
and
\eqref{eq:Casimir11}
but with the first term in both expressions appearing only once.

Some more invariant quadratic generators obviously
include quadratic combinations of the central charges
\[
\gen{C}^2=
\varepsilon_{\mathfrak{bc}}
\varepsilon_{\mathfrak{da}}
\gen{C}^{\mathfrak{ab}}
\gen{C}^{\mathfrak{cd}},
\qquad
\hat{\gen{C}}^2=
\varepsilon_{\mathfrak{bc}}
\varepsilon_{\mathfrak{da}}
\hat{\gen{C}}^{\mathfrak{ab}}
\hat{\gen{C}}^{\mathfrak{cd}},
\qquad
\hat{\gen{D}}^2.
\]
%

\subsection{Triplet of Cubic $\alg{psu}(1,1|2)$ Invariants}
\label{sec:cubic}

Curiously, there exist three cubic 
$\alg{psu}(1,1|2)$ invariants
$(\gen{J}^3){}^{\mathfrak{ab}}=(\gen{J}^3){}^{\mathfrak{ba}}$
for the algebra without central extensions, $\gen{C}^{\mathfrak{ab}}=0$,
\<\label{eq:cubic}
(\gen{J}^3){}^{\mathfrak{ab}}\eq
\half \varepsilon_{ce}
\varepsilon_{dh}
\varepsilon_{\zeta \iota }
\gen{R}^{cd}
\comm{\gen{Q}^{e \zeta \mathfrak{a}}}
{\gen{Q}^{h \iota \mathfrak{b}}}
+
\half \varepsilon_{eh}
\varepsilon_{\gamma \zeta}
\varepsilon_{\delta \iota  }
\gen{J}^{\gamma \delta}
\comm{\gen{Q}^{e \zeta \mathfrak{a}}}
{\gen{Q}^{h \iota \mathfrak{b}}}
\nlnum\nonumber
+
\varepsilon_{de}
\varepsilon_{fc}
\gen{B}^{\mathfrak{ab}}
\gen{R}^{cd}
\gen{R}^{ef} 
-
\varepsilon_{\delta \epsilon}
\varepsilon_{\zeta \gamma}
\gen{B}^{\mathfrak{ab}}
\gen{J}^{\gamma \delta}
\gen{J}^{\epsilon \zeta }
-
\varepsilon_{cf}
\varepsilon_{\delta\eta}
\varepsilon_{\mathfrak{eh}}
\gen{B}^{\mathfrak{ab}}
\gen{Q}^{c\delta\mathfrak{e}}
\gen{Q}^{f\eta\mathfrak{h}}.
\>
They transform as a triplet under $\gen{B}$,
and commute with the $\alg{psu}(1,1|2)$ algebra.
These cubic generators are important for 
the multiplet structure in the algebra with automorphism.
For a multiplet of the extended algebra,
the highest-weight states of $\alg{psu}(1,1|2)$ 
form a multiplet of $\alg{su}(2)$.
To move about in this multiplet, one cannot simply
use the $\alg{su}(2)$ generators $\gen{B}^{\mathfrak{ab}}$ 
because they do not commute with $\alg{psu}(1,1|2)$.
Instead, the cubic generators map between highest-weight
states of $\alg{psu}(1,1|2)$, i.e.~they can be understood
as $\alg{su}(2)$ ladder generators. 

\subsection{Algebra of Cubic Invariants}
\label{sec:cubicalg}

The cubic operators $(\gen{J}^3)^{\mathfrak{ab}}$
commute with all $\alg{psu}(1,1|2)$ generators, and
they transform as a triplet under the $\alg{su}(2)$ automorphism
\[
\comm{\gen{B}^{\mathfrak{ab}}}{ (\gen{J}^3)^{\mathfrak{cd}}}
=\varepsilon^{\mathfrak{cb}} (\gen{J}^3)^{\mathfrak{ad}}
-\varepsilon^{\mathfrak{ad}} (\gen{J}^3)^{\mathfrak{cb}}.
\]
It remains to be seen how they commute among themselves.

We first note that $(\gen{J}^3)^{\mathfrak{ab}}$ in \eqref{eq:cubic} contains
the quadratic Casimir $\gen{J}^2$ in \eqref{eq:Casimir112} 
(with $\gen{C}^{\mathfrak{ab}}=0$)
multiplied by the $\alg{su}(2)$ generator $\gen{B}^{\mathfrak{ab}}$.
We can thus split it up into two parts
\[
(\gen{J}^3){}^{\mathfrak{ab}} = (\tilde{\gen{J}}^3)^{\mathfrak{ab}}  
+ \gen{J}^2 \gen{B}^{\mathfrak{ab}}
\]
with the remainder 
\<\label{eq:cubicremainder}
(\tilde{\gen{J}}^3)^{\mathfrak{ab}} \eq  \half \varepsilon_{ce}
\varepsilon_{dh}
\varepsilon_{\zeta \iota }
\gen{R}^{cd}
\comm{\gen{Q}^{e \zeta \mathfrak{a}}}
{\gen{Q}^{h \iota \mathfrak{b}}}
+
\half \varepsilon_{eh}
\varepsilon_{\gamma \zeta}
\varepsilon_{\delta \iota  }
\gen{J}^{\gamma \delta}
\comm{\gen{Q}^{e \zeta \mathfrak{a}}}
{\gen{Q}^{h \iota \mathfrak{b}}}.
\>
Now, $(\gen{J}^3){}^{\mathfrak{ab}}$ commutes with ordinary 
$\alg{psu}(1,1|2)$ generators, and the $\tilde{\gen{J}}^3$
 are products of ordinary $\alg{psu}(1,1|2)$ generators only. 
Therefore, the commutator of two nonidentical $\gen{J}^3$ generators 
yields simply a product of the quadratic Casimir and a $\gen{J}^3$,
\[ \label{eq:J3level2}
 \comm{(\gen{J}^3)^{\mathfrak{ab}}}{ (\gen{J}^3)^{\mathfrak{cd}}}
= \varepsilon^{\mathfrak{cb}} \gen{J}^2(\gen{J}^3)^{\mathfrak{ad}}
 - \varepsilon^{\mathfrak{ad}} \gen{J}^2 (\gen{J}^3)^{\mathfrak{cb}}.
\]

From this, it is straightforward to obtain the entire algebra generated 
by the cubic invariants. Define 
\[ \label{eq:defineJ3n}
(\gen{J}_0^3)^{\mathfrak{ab}}=\gen{B}^{\mathfrak{ab}}\qquad\mbox{and}\qquad
(\gen{J}_n^3)^{\mathfrak{ab}}= \left(\gen{J}^2\right)^{n-1} (\gen{J}^3)^{\mathfrak{ab}}, \quad n \geq 1.
\]
It only takes a short computation to show that 
these $\gen{J}_n^3$ satisfy 
a loop algebra
(the same algebra as the $\yang_n$ in \secref{sec:loop})
\[\label{eq:J3alg}
\comm{(\gen{J}_m^3)^{\mathfrak{ab}}}{(\gen{J}_n^3)^{\mathfrak{cd}}}
=
\varepsilon^{\mathfrak{cb}}(\gen{J}_{m+n}^3)^{\mathfrak{ad}}
-\varepsilon^{\mathfrak{ad}}(\gen{J}_{m+n}^3)^{\mathfrak{cb}}.
\]
For $n$ or $m$ equal to 0, this algebra is satisfied since 
the quadratic Casimir commutes even with $\gen{B}^{\mathfrak{ab}}$. 
Assuming $n$ and $m$ are greater than 0, 
we substitute the definition \eqref{eq:defineJ3n} to obtain
\<
\comm{(\gen{J}_m^3)^{\mathfrak{ab}}}{(\gen{J}_n^3)^{\mathfrak{cd}}}
\eq \comm{\left(\gen{J}^2\right)^{m-1} (\gen{J}^3)^{\mathfrak{ab}}}{\left(\gen{J}^2\right)^{n-1} (\gen{J}^3)^{\mathfrak{cd}}}
\nln
\eq \left(\gen{J}^2\right)^{n+m-2}  \comm{(\gen{J}^3)^{\mathfrak{ab}}}{ (\gen{J}^3)^{\mathfrak{cd}}}
\nln
\eq \varepsilon^{\mathfrak{cb}} \left(\gen{J}^2\right)^{n+m-1} (\gen{J}^3)^{\mathfrak{ad}} - \varepsilon^{\mathfrak{ad}} \left(\gen{J}^2\right)^{n+m-1} (\gen{J}^3)^{\mathfrak{cb}}
\nln
\eq \varepsilon^{\mathfrak{cb}}  (\gen{J}^3_{n+m})^{\mathfrak{ad}} - \varepsilon^{\mathfrak{ad}}  (\gen{J}^3_{n+m})^{\mathfrak{cb}},
\>
as required. We used the vanishing commutator between $\gen{J}^2$ 
and $(\gen{J}^3)^{\mathfrak{ab}}$, and \eqref{eq:J3level2}. 
It is interesting that the role of the quadratic 
Casimir operator here resembles that of $\hat{\gen{D}}$ 
in the $\yang$ algebra for cyclic states above \eqref{eq:yalg}.

\subsection{Representation of the Algebra}

Let us understand the representations of the above loop algebra
generated by $(\gen{J}^3){}^{\mathfrak{ab}}$,
cf.~\secref{sec:rep}. We act with $({\gen{J}}^3_n)^{<>}$ 
on a $\alg{su}(2)\ltimes\alg{psu}(1,1|2)$ highest-weight state $\state{\Psi}$ 
and find
\<
({\gen{J}}^3_n)^{<>}\state{\Psi}
\eq
(\gen{J}^2)^{n-1} (\tilde{\gen{J}}^3)^{<>}\state{\Psi}
+
(\gen{J}^2)^n \gen{B}^{<>} \state{\Psi}
\nln\eq
x^{n-1} (\tilde{\gen{J}}^3)^{<>}\state{\Psi}
+
x^n \gen{B}^{<>} \state{\Psi},
\>
where $x$ is the eigenvalue of the quadratic Casimir $\gen{J}^2$ on $\state{\Psi}$.
Now it turns out that $(\tilde{\gen{J}}^3)^{<>} \state{\Psi} =0$,
and consequently
\[
({\gen{J}}^3_n)^{<>}\state{\Psi}
=
x^n \gen{B}^{<>} \state{\Psi}.
\]
Therefore, the representation of the loop algebra of $\gen{J}^3_n$
is an evaluation representation with evaluation parameter $x$. 
In the case of a $(m+1)$-dimensional $\alg{su}(2)$ multiplet
of $\alg{psu}(1,1|2)$ representations, 
the highest weight is realized as a symmetric tensor product of $m$ fundamental
evaluation representations with equal evaluation parameters $x$
\[
\state{<,x}\otimes\state{<,x}\otimes\ldots \otimes\state{<,x}.
\]

\subsection{A One-Parameter Deformation of the Loop Algebra}
\label{sec:cubicalgebra}

Assuming that the $\yang^{\mathfrak{ab}}$ satisfy a loop algebra
as explained in \secref{sec:loop}, 
there is actually a one-parameter generalization 
of these generators using the $(\gen{J}^3)^{\mathfrak{ab}}$. 
The same subalgebra of the $\alg{su}(2)$ loop algebra 
is generated by 
\[
\hat{\yang}^{\mathfrak{ab}} = \yang^{\mathfrak{ab}} + \alpha \, (\gen{J}^3)^{\mathfrak{ab}}
\]
for any constant $\alpha$.

Let us present the full deformation of the $\yang_n^{\mathfrak{ab}}$, 
which we label $\hat{\yang}_n^{\mathfrak{ab}}$. 
We parameterize the deformation with $\alpha$. 
$\hat{\yang}_0^{\mathfrak{ab}}$ is still given by $\gen{B}^{\mathfrak{ab}}$, 
but all other generators become
\[ \label{eq:defineyhat}
\hat{\yang}_n^{\mathfrak{ab}}=\alpha^n (\gen{J}_n^3)^{\mathfrak{ab}} 
+ \sum_{m=0}^{n-1} \alpha^m \binom{n}{m}  \left(\gen{J}^2\right)^m \yang_{n-m}^{\mathfrak{ab}}, \quad n \geq 1.
\]
Again, the algebra relations take the same form,
\[\label{eq:yhatalg}
\comm{\hat{\yang}_m^{\mathfrak{ab}}}{\hat{\yang}_n^{\mathfrak{cd}}}
=
\varepsilon^{\mathfrak{cb}}\hat{\yang}_{m+n}^{\mathfrak{ad}}
-\varepsilon^{\mathfrak{ad}}\hat{\yang}_{m+n}^{\mathfrak{cb}}.
\]
The commutators with $n$ or $m$ equal to zero are again 
satisfied because $\gen{J}^2$ commutes with $\gen{B}^{\mathfrak{ab}}$. 
In order to check the relations with $m=1$, 
we need the commutators between the $(\gen{J}^3_n)^{\mathfrak{ab}}$ 
and the $\yang_1^{\mathfrak{ab}}$.
The vanishing commutator between the $\yang$ 
and the ordinary $\alg{psu}(1,1|2)$ generators implies for all $n \geq 0$
\begin{equation} \label{eq:yJ3comm}
   \comm{\yang_1^{\mathfrak{ab}}}{ (\gen{J}^3_n)^{\mathfrak{cd}}}  =   \left( \gen{J}^2 \right)^{n} \left( \varepsilon^{\mathfrak{cb}} \yang_{1}^{\mathfrak{ad}}  -   \varepsilon^{\mathfrak{ad}} \yang_{1}^{\mathfrak{cb}} \right).
  \end{equation}
Now we expand the left side of the relations \eqref{eq:yhatalg} 
with $m=1$ using \eqref{eq:defineyhat}. 
We simplify using \eqref{eq:yJ3comm} 
and the algebras of the $(\gen{J}^3_n)^{\mathfrak{ab}}$ and of the $\yang_n^{\mathfrak{ab}}$,
\<
\comm{\hat{\yang}_1^{\mathfrak{ab}}}{\hat{\yang}_n^{\mathfrak{cd}}} \eq \alpha^n \comm{\yang_1^{\mathfrak{ab}}}{ (\gen{J}_n^3)^{\mathfrak{cd}}}+  \alpha^{n+1} \comm{(\gen{J}^3)^{\mathfrak{ab}}}{ (\gen{J}_n^3)^{\mathfrak{cd}}} 
\nl
+ \sum_{m=0}^{n-1} \alpha^m \binom{n}{m}  \left( \comm{\yang_1^{\mathfrak{ab}}}{ \left(\gen{J}^2\right)^m \yang_{n-m}^{\mathfrak{cd}}}  + \alpha \comm{(\gen{J}^3)^{\mathfrak{ab}}}{ \left(\gen{J}^2\right)^m \yang_{n-m}^{\mathfrak{cd}}} \right)
\nln
\eq \alpha^n \varepsilon^{\mathfrak{cb}} \left(\gen{J}^2\right)^n \yang_1^{\mathfrak{ad}}- \alpha^n \varepsilon^{\mathfrak{ad}} \left(\gen{J}^2\right)^n \yang_1^{\mathfrak{cb}}+ \alpha^{n+1} \varepsilon^{\mathfrak{cb}} (\gen{J}^3_{n+1})^{\mathfrak{ad}}-  \alpha^{n+1} \varepsilon^{\mathfrak{ad}} (\gen{J}^3_{n+1})^{\mathfrak{cb}}
\nl
+ \sum_{m=0}^{n-1} \alpha^m \binom{n}{m}  \left(\gen{J}^2\right)^m \left( \varepsilon^{\mathfrak{cb}} \yang_{n+1-m}^{\mathfrak{ad}}  -   \varepsilon^{\mathfrak{ad}} \yang_{n+1-m}^{\mathfrak{cb}} \right)
\nl
+\sum_{m=0}^{n-1} \alpha^{m+1} \binom{n}{m} \left(\gen{J}^2\right)^{m+1} \left( \varepsilon^{\mathfrak{cb}} \yang_{n-m}^{\mathfrak{ad}}  -   \varepsilon^{\mathfrak{ad}} \yang_{n-m}^{\mathfrak{cb}} \right)
\nln
\eq \alpha^{n+1} \varepsilon^{\mathfrak{cb}} (\gen{J}^3_{n+1})^{\mathfrak{ad}}-  \alpha^{n+1} \varepsilon^{\mathfrak{ad}} (\gen{J}^3_{n+1})^{\mathfrak{cb}} 
\nl
+  \sum_{m=0}^{n} \alpha^m \binom{n+1}{m}  \left(\gen{J}^2\right)^m \left( \varepsilon^{\mathfrak{cb}} \yang_{n+1-m}^{\mathfrak{ad}}  -   \varepsilon^{\mathfrak{ad}} \yang_{n+1-m}^{\mathfrak{cb}} \right) 
\nln
\eq \varepsilon^{\mathfrak{cb}}\hat{\yang}_{n+1}^{\mathfrak{ad}}
-\varepsilon^{\mathfrak{ad}}\hat{\yang}_{n+1}^{\mathfrak{cb}}.
\>
We combined terms to reach the second-to-last expression, 
and substituted the definition \eqref{eq:defineyhat} for the last line. 
The calculation proceeds in parallel for \eqref{eq:yhatalg} with $n=1$. 
Since this algebra's Serre relations are the level three equations, 
which have $n$ or $m$ equal to one, it follows that \eqref{eq:yhatalg} is satisfied.

\section{Symmetries of the Bilocal Generators} 
\label{app:biproof}

In this appendix, we prove that the $\yang^{\mathfrak{ab}}$ 
commute with the $\alg{psu}(1,1|2)$ generators, 
including the one-loop dilatation generator. 
The proofs can be modified straightforwardly 
to show the same for $\yangsing$. 

Again, since we work only at leading order the  $\alg{psu}(1,1|2)$ generators
  $\gen{Q},\gen{J}$ are truncated at $\order{g^0}$, and the $\alg{psu}(1|1)^2$
generators $\hat{\gen{Q}},\hat{\gen{S}}$ only act with $\hat{\gen{Q}}_{(1)},\hat{\gen{S}}_{(1)}$.
\subsection{Invariance under $\alg{psu}(1,1|2)$}
\label{sec:proof112}

We now prove that $\yang^{\mathfrak{ab}}$ commutes with 
the classical $\alg{psu}(1,1|2)$ generators. 
It is sufficient to prove that the commutators with the $\gen{Q}$ 
vanish since the $\gen{Q}$ generate the complete algebra. 
Furthermore, using $\gen{B}$ symmetry, 
it is sufficient to prove this for $\yang^{\mathfrak{<<}}$. 
Now, $\gen{Q}^{a\beta<}$ commute exactly with the $\hat{\gen{Q}}^<$ 
and $\hat{\gen{S}}^<$, so it is clear that these commutators vanish. 
However, it is nontrivial to show that the $\gen{Q}^{a+>}$ 
commute with $\yang^{<<}$, since they only commute with $\hat{\gen{Q}}^<$ up to a 
gauge transformation
\[
\acomm{\gen{Q}^{a+>}}{\hat{\gen{Q}}^<} \state{X} 
=  \varepsilon^{ab} \state{X \phi^{(0)}_b}
- \varepsilon^{ab} \state{\phi^{(0)}_b X}= \check{Z}^a(2) - \check{Z}^a(1).
\label{eq:gaugetransformation} \]
Here we use the notation $\check{Z}^a(i) $ 
for the insertion of a bosonic field at a new site between 
the original sites $i$ and $i+1$. 
It will be useful to note that we can use $\shift$ to change the 
site indices of any generator that acts on site $i$ 
and any number of following sites,
\[
X (i+1 \ldots) = \shift X(i \ldots) \shift^{-1}.
\label{eq:shiftsite} \]
We are now ready to check the commutator directly. 
We use that the $\gen{Q}^{a+>}$ still commute exactly 
with $\hat{\gen{S}}^<$ and  apply \eqref{eq:gaugetransformation} and \eqref{eq:shiftsite},
\<
\comm{\gen{Q}^{a+>}}{\yang^{\mathfrak{<<}}} \eq  \sum_{j=0}^{L-1} \sum_{i=0}^{L+1}(1 - \frac{\delta_{i,\,0}}{2} - \frac{\delta_{i,\, L+1}}{2}) \shift^{j-i} \, \hat{\gen{S}}^{<}(1,2)\,  \shift^i \, (\check{Z}^{a} (1) - \check{Z}^{a} (2)) \shift^{-j} \nln
\eq   \sum_{j=0}^{L-1} \sum_{i=0}^{L+1} (1 - \frac{\delta_{i,\,0}}{2} - \frac{\delta_{i,\, L+1}}{2}) \shift^{j-i} \, \hat{\gen{S}}^{<}(1,2)\,  \shift^i \, \check{Z}^{a} (1) \shift^{-j} 
   \nln
\earel{-}
 \sum_{j=0}^{L-1} \sum_{i=0}^{L+1} (1 - \frac{\delta_{i,\,0}}{2} - \frac{\delta_{i,\, L+1}}{2}) \shift^{(j+1)-(i+1)} \, \hat{\gen{S}}^{<}(1,2)\,  \shift^{i+1} \, \check{Z}^{a} (1) \shift^{-(j+1)} 
 \nln
\eq \sum_{j=0}^{L-1} \sum_{i=0}^{L+1} (1 - \frac{\delta_{i,\,0}}{2} - \frac{\delta_{i,\, L+1}}{2}) \shift^{j-i} \, \hat{\gen{S}}^{<}(1,2)\,  \shift^i \, \check{Z}^{a} (1) \shift^{-j}  
 \nln
\earel{-}   \sum_{j=1}^{L} \sum_{i=1}^{L+2} (1 - \frac{\delta_{i,\,1}}{2} - \frac{\delta_{i,\, L+2}}{2}) \shift^{j-i} \, \hat{\gen{S}}^{<}(1,2)\,  \shift^{i} \, \check{Z}^{a} (1) \shift^{-j}
\>
We shifted summation variables to obtain the last line, 
$i \rightarrow (i+1)$ and \\
$j \rightarrow (j+1)$.  
Since the chain is of length $L$ initially and after the application 
of the commutator, $j=L$ is equivalent to $j=0$.
 Now we can combine the two lines (being careful with the different ranges for $i$) and simplify,
\<
\comm{\gen{Q}^{a+>}}{\yang^{\mathfrak{<<}}} \eq 
\sum_{j=0}^{L-1} \sum_{i=0}^{L+2} \Big[ \big((1 - \frac{\delta_{i,\,0}}{2} - \frac{\delta_{i,\,L+1}}{2}  - \delta_{i,\,L+2})-(1 - \frac{\delta_{i,\,1}}{2} - \frac{\delta_{i,\,L+2}}{2}  - \delta_{i,\,0})\big) 
\nln
&& \quad \times   \, \,   \shift^{j-i} \, \hat{\gen{S}}^{<}(1,2)\,  \shift^i \, \check{Z}^{a} (1) \shift^{-j} \Big] \nln
\eq  
\frac{1}{2} \sum_{j=0}^{L-1} ( \shift^{j-1} \, \hat{\gen{S}}^{<}(1,2)\,  \shift \, \check{Z}^{a} (1) \shift^{-j} + \shift^{j} \, \hat{\gen{S}}^{<}(1,2) \, \check{Z}^{a} (1) \shift^{-j} 
\nln
\earel{-} \shift^{j-2} \, \hat{\gen{S}}^{<}(1,2)\,  \shift \, \check{Z}^{a} (1) \shift^{-j}- \shift^{j-1} \, \hat{\gen{S}}^{<}(1,2) \, \check{Z}^{a} (1) \shift^{-j}) \nln
\eq 
\frac{1}{2} (1 - \shift^{-1})  \sum_{j=0}^{L-1} \shift^j (\shift^{-1} \, \hat{\gen{S}}^{1}(1,2)\,  \shift \, \check{Z}^{a} (1)+  \hat{\gen{S}}^{<}(1,2)\,   \check{Z}^{a} (1)  ) \shift^{-j}.
\>
To reach the middle expressions, we used that the length 
of the chain is $L+1$ after $\check{Z}^a$ acts. 
The expression in parenthesis inside the sum in the 
last line gives a chain derivative by parity. 
To see this, we write the chain with site $0=L$ first:
\<
(\shift^{-1} \, \hat{\gen{S}}^{<}(1,2)\,  \shift \, \check{Z}^{a} (1)+  \hat{\gen{S}}^{<}(1,2)\,   \check{Z}^{a} (1)) \state{Y_0 \, Y_1 \, Y_2 \ldots} \eq \nln 
\frac{\varepsilon^{ab}}{2} ( \gen{S}^<(0,1) \state{Y_{0} \phi^{(0)}_b \, Y_1 \, Y_2 \ldots } + \gen{S}^<(1,2)  \state{Y_0 \, \phi^{(0)}_b \, Y_1 \, Y_2 \ldots}) \eq \nln
\frac{\varepsilon^{ab}}{2} ( -\gen{S}^<(0,1) \state{\phi^{(0)}_b Y_{0}  \, Y_1 \, Y_2 \ldots } + \gen{S}^<(1,2)  \state{Y_0 \, \phi^{(0)}_b \, Y_1 \, Y_2 \ldots}).
\>
We used parity to reach the last line. Since this 
term acts homogeneously on the chain, the first and second terms cancel.

The proof for the $\gen{Q}^{a->}$ is similar.  
They only commute with $\hat{\gen{S}}^<$ up to the gauge transformations 
\<
\acomm{\gen{Q}^{a->}}{\hat{\gen{S}}^<} \state{X \phi^{(0)}_b } \eq -\delta^{a}_b \state{ X }= -\hat{Z}^a(2), \nln
\acomm{\gen{Q}^{a->}}{\hat{\gen{S}}^<} \state{\phi^{(0)}_b X } \eq \delta^{a}_b \state{ X }=  \hat{Z}^a(1). 
\>
Here we have defined $\hat{Z}^a(i)$.  
Since the $\gen{Q}^{a->}$  commute exactly with 
the $\hat{\gen{Q}}^<$, again using \eqref{eq:shiftsite} 
 to shift site indices we find
\<
\comm{\gen{Q}^{a->}}{\yang^{\mathfrak{<<}}} \eq  \sum_{j=0}^{L-1} \sum_{i=0}^{L+1}(1 - \frac{\delta_{i,\,0}}{2} - \frac{\delta_{i,\, L+1}}{2}) \shift^{j-i} \, (\hat{Z}^{a}(1) - \hat{Z}^{a}(2))\,  \shift^i \, \hat{\gen{Q}}^{<} (1)  \shift^{-j} \nln
\eq 
\frac{1}{2} (\shift^{-1}-1)  \sum_{j=0}^{L-1} \shift^j (\shift \,  \hat{Z}^{a} (1) \, \shift^{-1} \, \hat{\gen{Q}}^{<}(1) +  \hat{Z}^{a} (1) \, \hat{\gen{Q}}^{<}(1)  ) \shift^{-j}.
\>
Again, the term in parenthesis is a chain derivative by parity. 
This completes the proof that the $\gen{Q}$ commute with $\yang^{<<}$. 
It follows by $\gen{B}$ and $\alg{psu}(1,1|2)$ 
symmetry that the $\yang^{\mathfrak{ab}}$ commute 
with all of the classical $\alg{psu}(1,1|2)$ generators. 

It is clear from the above proof that $\yangsing$ \eqref{eq:nonlocalgen2}
also commutes with the classical $\alg{psu}(1,1|2)$ generators, 
since the bilocal product of $\hat{\gen{S}}^<$ and $\hat{\gen{Q}}^>$ 
(or $\hat{\gen{S}}^>$ and $\hat{\gen{Q}}^<$) by itself commutes.

\subsection{Conservation}
\label{sec:proofham}

To prove that the $\yang$ commute with the Hamiltonian $\ham$, or with $\hat{\gen{D}}=\half \, \ham$, 
we first need to consider how the $\alg{psu}(1|1)^2$ 
generators commute with the Hamiltonian. Locally, we have
\<
\hat{\gen{D}} \eq \acomm{\hat{\gen{Q}}^<} {\hat{\gen{S}}^>} + \, \text{chain derivative}, \nln
 \eq - \acomm{\hat{\gen{Q}}^>} {\hat{\gen{S}}^<} + \, \text{chain derivative}, \nln
 \eq \half \delta \gen{D}_2 + \, \text{chain derivative}.
\label{eq:gaugeequivalentD}
 \>
Here ``locally'' refers to the interactions that are summed 
over the length of the chain. For instance, the local expression 
for the one-loop commutators expand as one-site to one-site and 
two-site to two-site interactions,
\[
\acomm{\hat{\gen{Q}}^{\mathfrak{a}}} {\hat{\gen{S}}^{\mathfrak{b}}} 
=  (\hat{\gen{S}}^{\mathfrak{b}}(1,2)   \hat{\gen{Q}}^{\mathfrak{a}}(1))
 + ( \hat{\gen{Q}}^{\mathfrak{a}}(1) \hat{\gen{S}}^{\mathfrak{b}}(1,2)
 + \hat{\gen{S}}^{\mathfrak{b}}(2,3)   \hat{\gen{Q}}^{\mathfrak{a}}(1) 
 +\hat{\gen{S}}^{\mathfrak{b}}(1,2)   \hat{\gen{Q}}^{\mathfrak{a}}(2)).
\label{eq:localexpansion} \]
The term inside the first parenthesis is one-site to one-site,
 and the remaining terms are two-site to two-site. 
A chain derivative summed over the length of a periodic chain gives zero, 
so when we commute $\yang^{\mathfrak{ab}}$ with the Hamiltonian, 
we can use any of the equivalent forms in \eqref{eq:gaugeequivalentD}
 as long as each one acts homogeneously on the chain. 
We will use this freedom to always commute any $\alg{psu}(1,1)^2$ 
generator with the commutator in \eqref{eq:gaugeequivalentD}
that involves the same generator. Therefore, it will be convenient to define
\<
\gen{D}_L \eq  \acomm{\hat{\gen{Q}}^<} {\hat{\gen{S}}^>} \nln
\gen{D}_R \eq - \acomm{\hat{\gen{Q}}^>} {\hat{\gen{S}}^<}.
\label{eq:defineDs} \>
Furthermore, the $\gen{D}_L$ and $\gen{D}_R $ split into local 
one-site to one-site and two-site to two-site interactions \eqref{eq:localexpansion}.
Then we have the exact local equalities only involving the 
two-site to two-site interactions of $\gen{D}_L$ and $\gen{D}_R $,
\<
\comm{\hat{\gen{Q}}^<(i)}{\gen{D}_L} \eq  \hat{\gen{q}}^<(i-1, i) - \hat{\gen{q}}^<(i, i+1), \nln
\comm{\hat{\gen{Q}}^>(i)}{\gen{D}_R} \eq  \hat{\gen{q}}^>(i-1, i) - \hat{\gen{q}}^>(i, i+1), \nln
\hat{\gen{q}}^<(i-1, i) \eq   \hat{\gen{Q}}^<(i) \gen{D}_L(i-1,i)-   \gen{D}_L(i-1,i) \hat{\gen{Q}}^<(i) \nln
\hat{\gen{q}}^>(i-1, i) \eq   \hat{\gen{Q}}^>(i) \gen{D}_R(i-1,i)-   \gen{D}_R(i-1,i) \hat{\gen{Q}}^>(i).
\>
Note that the $\hat{\gen{q}}(i, i+1)$ have two-site to three-site interaction,
 with final sites $(i, i+1, i+2)$, and that their explicit forms in terms
of interactions are not needed. These equalities can be shown easily by 
expanding $\gen{D}_L$ and $\gen{D}_R$ and using
 the fact that $(\hat{\gen{Q}}^{\mathfrak{a}})^2=0$ 
is even satisfied on a one-site chain:
\[
(\hat{\gen{Q}}^{\mathfrak{a}}(1) + \hat{\gen{Q}}^{\mathfrak{a}}(2))\hat{\gen{Q}}^{\mathfrak{a}}(1)=0 \quad \text{(no sum).}
\label{eq:onesiteQ^2} \]
Similarly, we have
 \<
\comm{\hat{\gen{S}}^<(i, i+1)}{\gen{D}_R} \eq  \hat{\gen{s}}^<(i-1, i, i+1) - \hat{\gen{s}}^<(i, i+1, i+2), \nln
\comm{\hat{\gen{S}}^>(i, i+1)}{\gen{D}_L} \eq  \hat{\gen{s}}^>(i-1, i, i+1) - \hat{\gen{s}}^>(i, i+1, i+2), \nln
\hat{\gen{s}}^<(i-1, i, i+1) \eq  \hat{\gen{S}}^<(i, i+1) \gen{D}_R(i-1,i)-   \gen{D}_R(i-1,i) \hat{\gen{S}}^<(i, i+1) \nln
\hat{\gen{s}}^>(i-1, i, i+1) \eq   \hat{\gen{S}}^>(i,i+1) \gen{D}_L(i-1,i)-   \gen{D}_L(i-1,i) \hat{\gen{S}}^>(i, i+1).
\>
 The $\hat{\gen{s}}(i, i+1, i+2)$ have three-site to two-site 
interactions, with final sites $(i, i+1)$, and again we do not need their explicit forms.
Now, using these commutation relations, and the identities 
that follow from \eqref{eq:shiftsite}
\<
\hat{\gen{q}}^\mathfrak{a}(i-1, i) 
\eq \shift^{-1}\, \hat{\gen{q}}^\mathfrak{a}(i, i+1) \, \shift, \nln
\hat{\gen{s}}^\mathfrak{a}(i-1, i, i+1) 
\eq \shift^{-1}\, \hat{\gen{s}}^\mathfrak{a}(i, i+1, i+2)\,  \shift,
\>
we find 
\<
\comm{\hat{\gen{D}}}{\yang^{\mathfrak{ab}}} 
\eq \sum_{j=0}^{L-1} \sum_{i=0}^{L+1}(1 - \half\delta_{i,0} - \half\delta_{i,L+1}) \shift^{j-i} \, (\hat{\gen{s}}^{\mathfrak{a}}(0, 1,2) - \hat{\gen{s}}^{\mathfrak{a}}(1, 2, 3))\,  \shift^i \, \hat{\gen{Q}}^{\mathfrak{b}} (1)  \shift^{-j} 
\nl
+
\sum_{j=0}^{L-1} \sum_{i=0}^{L+1}(1 - \half\delta_{i,0} - \half\delta_{i,L+1}) \shift^{j-i} \,  \hat{\gen{S}}^{\mathfrak{a}} (1, 2)   \shift^i \, (\hat{\gen{q}}^{\mathfrak{b}}(0, 1) - \hat{\gen{q}}^{\mathfrak{b}}(1, 2))\, \shift^{-j} 
\nln
\eq
-\half (1 - \shift^{-1}) \sum_{j=0}^{L-1} \shift^{j} \, ( \hat{\gen{s}}^{\mathfrak{a}}(1, 2,3) \, \hat{\gen{Q}}^{\mathfrak{b}} (1) + \shift^{-1} \, \hat{\gen{s}}^{\mathfrak{a}}(1, 2,3)\,  \shift \, \hat{\gen{Q}}^{\mathfrak{b}} (1) )   \shift^{-j} 
\nl
+
\half (1 - \shift^{-1}) \sum_{j=0}^{L-1} \shift^{j} \, ( \hat{\gen{S}}^{\mathfrak{a}}(1, 2) \, \hat{\gen{q}}^{\mathfrak{b}} (1,2) + \shift \, \hat{\gen{S}}^{\mathfrak{a}}(1, 2)\,  \shift^{-1} \, \hat{\gen{q}}^{\mathfrak{b}} (1,2) )   \shift^{-j}.
\nl
\>
To complete the proof, we will now show that this vanishes
 since it is a homogeneous sum of a chain derivative. 
Equivalently,
\[
 \hat{\gen{S}}^{\mathfrak{a}}(1, 2) \, \hat{\gen{q}}^{\mathfrak{b}} (1,2) + \shift \, \hat{\gen{S}}^{\mathfrak{a}}(1, 2)\,  \shift^{-1} \, \hat{\gen{q}}^{\mathfrak{b}} (1,2) - \hat{\gen{s}}^{\mathfrak{a}}(1, 2,3) \, \hat{\gen{Q}}^{\mathfrak{b}} (1) - \shift^{-1} \, \hat{\gen{s}}^{\mathfrak{a}}(1, 2,3)\,  \shift \, \hat{\gen{Q}}^{\mathfrak{b}} (1), \label{YHcommutatorlocal}
 \]
acts as a chain derivative on sites 1 and 2. 

First we simplify the first term. For simplicity,
we consider the $(<<)$ component.
 By definition and using the two-site to two-site interactions 
of the defining commutator of $\gen{D}_L$ \eqref{eq:defineDs}, 
we have
\<
 \hat{\gen{q}}^<(1,2) \eq    \hat{\gen{Q}}^<(2) \gen{D}_L(1,2)-    \gen{D}_L(1,2) \hat{\gen{Q}}^<(2) \nln
\eq  \hat{\gen{Q}}^<(2)  \hat{\gen{Q}}^<(1) \hat{\gen{S}}^>(1,2)+  \hat{\gen{Q}}^<(2)  \hat{\gen{S}}^> (1,2) \hat{\gen{Q}}^<(2) +  \hat{\gen{Q}}^<(2)  \hat{\gen{S}}^> (2,3) \hat{\gen{Q}}^<(1) \nln
\earel{-}
\hat{\gen{Q}}^<(1) \hat{\gen{S}}^>(1,2)  \hat{\gen{Q}}^<(2)-   \hat{\gen{S}}^> (1,2) \hat{\gen{Q}}^<(2)  \hat{\gen{Q}}^<(2) -   \hat{\gen{S}}^> (2,3) \hat{\gen{Q}}^<(1) \hat{\gen{Q}}^<(2).
\nonumber\\\earel{}
\>
Now, in the second term of the last expression (on the second-to-last line), 
we can switch the order of  $\hat{\gen{Q}}^<(2)$  and $\hat{\gen{S}}^< (1,2)$ 
(with a minus sign) since these two operators do not act on any shared sites, 
but being careful with site indices, we must use $\hat{\gen{Q}}^<(3)$ instead. 
Then, by the identity \eqref{eq:onesiteQ^2} that $\hat{\gen{Q}}^2=0$ even on one site, 
we find that the second term and the fifth term cancel, 
and we are left with the simpler expression
\<
\hat{\gen{Q}}^<(2)  \hat{\gen{Q}}^<(1) \hat{\gen{S}}^>(1,2) +  \hat{\gen{Q}}^<(2)  \hat{\gen{S}}^> (2,3) \hat{\gen{Q}}^<(1) \nln
- \hat{\gen{Q}}^<(1) \hat{\gen{S}}^>(1,2)  \hat{\gen{Q}}^<(2)-    \hat{\gen{S}}^> (2,3) \hat{\gen{Q}}^<(1) \hat{\gen{Q}}^<(2). \label{simplifiedqhat}
\>
Now the first two terms of \eqref{YHcommutatorlocal} can be written as 
\[
\hat{\gen{S}}^{<}(1, 2) \, \hat{\gen{q}}^{<} (1,2)+ \hat{\gen{S}}^{<}(2, 3) \, \hat{\gen{q}}^{<} (1,2). \label{firsttwoterms}
\]
The contributions from the first term of \eqref{simplifiedqhat}
cancel using \eqref{eq:onesiteQ^2} and the identity%
\footnote{This identity can be proved without too much difficulty. 
The second term is minus the parity image of the first term, 
so one just needs to check that the first term is parity even. 
This can be done with a short computation because,  
by $\gen{B}$ charge conservation, the only possible interactions are of the form
  (suppressing derivatives) $ \state{\psi_{<}} \rightarrow \state{\psi_{>} \psi_{>}}$.}
\[
(\hat{\gen{S}}^{<}(1, 2) \, \hat{\gen{Q}}^<(1) -\hat{\gen{S}}^{<}(2, 3) \, \hat{\gen{Q}}^<(2)) \hat{\gen{Q}}^<(1) =0
\]
So we are left with the following six terms for \eqref{firsttwoterms}
(the first two terms of \eqref{YHcommutatorlocal})
\<
\hat{\gen{S}}^{<}(1, 2) \hat{\gen{Q}}^<(2)  \hat{\gen{S}}^> (2,3) \hat{\gen{Q}}^<(1) - \hat{\gen{S}}^{<}(1, 2) \hat{\gen{Q}}^<(1) \hat{\gen{S}}^>(1,2)  \hat{\gen{Q}}^<(2) \earel{-} \nln
\hat{\gen{S}}^{<}(1, 2) \hat{\gen{S}}^> (2,3) \hat{\gen{Q}}^<(1) \hat{\gen{Q}}^<(2) + \hat{\gen{S}}^{<}(2, 3) \hat{\gen{Q}}^<(2)  \hat{\gen{S}}^> (2,3) \hat{\gen{Q}}^<(1) \earel{-} \nln
\hat{\gen{S}}^{<}(2, 3)  \hat{\gen{Q}}^<(1) \hat{\gen{S}}^>(1,2)  \hat{\gen{Q}}^<(2)-    \hat{\gen{S}}^{<}(2, 3)  \hat{\gen{S}}^> (2,3) \hat{\gen{Q}}^<(1) \hat{\gen{Q}}^<(2). \earel{} 
 \label{expandedfirsttwoterms} \>
Similar steps can be used for the last two terms of 
\eqref{YHcommutatorlocal}. We find
\<
\hat{\gen{S}}^{<}(2, 3) \hat{\gen{Q}}^>(1)  \hat{\gen{S}}^< (1,2) \hat{\gen{Q}}^<(1)+ \hat{\gen{S}}^{<}(2, 3) \hat{\gen{S}}^{<}(1, 2) \hat{\gen{Q}}^>(2)  \hat{\gen{Q}}^<(1) \earel{-} \nln
\hat{\gen{S}}^{<}(1, 2) \hat{\gen{Q}}^>(2)  \hat{\gen{S}}^< (2,3) \hat{\gen{Q}}^<(1) + \hat{\gen{S}}^{<}(2, 3) \hat{\gen{Q}}^>(1)  \hat{\gen{S}}^< (1,2) \hat{\gen{Q}}^<(2) \earel{+} \nln
\hat{\gen{S}}^{<}(2, 3) \hat{\gen{S}}^{<}(1, 2) \hat{\gen{Q}}^>(2)  \hat{\gen{Q}}^<(2) -     \hat{\gen{S}}^{<}(1, 2) \hat{\gen{Q}}^>(2)  \hat{\gen{S}}^< (2,3) \hat{\gen{Q}}^<(2). \earel{} 
 \label{expandedlasttwoterms} \>

Recall that we need to show that \eqref{YHcommutatorlocal}
is a chain derivative. \eqref{YHcommutatorlocal}
 is the sum of  \eqref{expandedfirsttwoterms} and \eqref{expandedlasttwoterms}. 
At this point, it is necessary to explicitly expand these terms 
as a sum of interactions. However, we can use discrete symmetries 
to greatly reduce the amount of computation. 
Under the discrete transformation $R$ that acts as
\[
R \state{\psi_{<}^{(n)}}=\state{\psi_{>}^{(n)}}, \quad R \state{\psi_{>}^{(n)}}=\state{\psi_{<}^{(n)}}, \quad R \state{\phi_1^{(n)}}=\state{\phi_1^{(n)}},  \quad R \state{\phi_2^{(n)}}=-\state{\phi_2^{(n)}},
\]
the supercharges transform as
\[
R \hat{\gen{Q}}^< R^{-1}=-\hat{\gen{Q}}^>, \quad  R \hat{\gen{S}}^< R^{-1}=\hat{\gen{S}}^>,
\]
as can be confirmed by examining the expressions
 for the  $\alg{psu}(1|1)^2$ generators \eqref{TransQhat}.
 Then under the combined operation
\[
X \rightarrow R X^\dagger R^{-1},
\]
\eqref{expandedfirsttwoterms} transforms into minus 
\eqref{expandedlasttwoterms} (term by term). Also,  
\eqref{expandedfirsttwoterms} and  \eqref{expandedlasttwoterms}
are both parity odd. Using these discrete symmetries, 
as well as $\gen{R}$ symmetry and conservation of $\gen{B}$ charge, 
one can infer the complete action of \eqref{YHcommutatorlocal}
by computing the following four types of interactions:
 \<
  \state{\phi_1^{(n)} \phi_2^{(m)}} \earel{\longrightarrow}
\sum_{k=0}^{n+m-1} g_1 (n,m,k) \state{\psi_{>}^{(k)} \psi_{>}^{(n+m-k-1)}},
  \nln
  \state{\phi_1^{(n)} \psi_<^{(m)}} \earel{\longrightarrow}
\sum_{k=0}^{n+m}  g_2 (n,m,k) \state{\phi_{1}^{(k)} \psi_{>}^{(n+m-k)}}
+   g_3 (n,m,k)  \state{\psi_{>}^{(k)} \phi_{1}^{(n+m-k)}} ,
  \nln
  \state{\psi_<^{(n)} \psi_<^{(m)}}\earel{\longrightarrow}
\sum_{k=0}^{n+m}  g_4 (n,m,k) \state{\psi_{<}^{(k)} \psi_{>}^{(n+m-k)} }.
  \>
Completing this still lengthy computation, and applying the known 
symmetries, we find that the $<<$ component of \eqref{YHcommutatorlocal}
 is given by the chain derivative  $X^{<<}(1) - X^{<<}(2)$, 
where the only nonvanishing action of $X^{<<}$ is 
\[
X^{<<} \state{\psi_<^{(n)}} = \frac{2}{(n+1)^2} \state{\psi_>^{(n)}}.
\]
Therefore, the $<<$ component of the commutator with the Hamiltonian 
vanishes on periodic states, and by $\gen{B}$ 
symmetry the $\yang^{\mathfrak{ab}}$ commute 
with $\ham$. 

Analogous steps to those above can be used to show that $\yangsing$ 
also commutes with the Hamiltonian. However, we have only computed 
(via \texttt{Mathematica}) the two-site to two-site interactions 
in this case up to five excitations.
 That computation was consistent with the commutator being a chain derivative, 
but another lengthy computation is needed to complete the proof in this case 
(the five-excitation computation is extremely strong evidence).

\bibliographystyle{nb}
\bibliography{SU112Ext}

\end{document}